\documentclass[aps,showpacs,eqsecnum,floatfix,nofootinbib]{revtex4}

\usepackage{amsmath}

\usepackage[dvips]{graphicx,color}

\begin{document}

\title{Subtractive renormalization of the chiral potentials up to next-to-next-to-leading
order in higher NN partial waves}
\author{C.-J.~Yang, Ch.~Elster, and D.~R.~Phillips}
\affiliation{Institute of Nuclear and Particle Physics, and Department of Physics and
Astronomy, Ohio University,\\
Athens, OH 45701}
\email{cjyang, elster, phillips@phy.ohiou.edu}
\date{\today}

\begin{abstract}
We develop a subtractive renormalization scheme to evaluate the P-wave NN scattering phase shifts 
using chiral effective theory potentials. This allows us to consider arbitrarily high 
cutoffs in the Lippmann-Schwinger equation (LSE). We employ NN potentials computed  up to 
next-to-next-to-leading order (NNLO) in chiral effective theory, using both dimensional regularization and 
spectral-function regularization. Our results obtained from the subtracted P-wave LSE
 show that renormalization of the 
NNLO potential can be achieved by using the generalized NN scattering lengths as input---an 
alternative to fitting the constant that multiplies the P-wave contact interaction in the chiral effective theory 
NN force. However, in order to obtain a reasonable fit to the NN data at NNLO the generalized scattering 
lengths must be varied away from the values extracted from the so-called high-precision potentials. 
We investigate how the generalized scattering lengths extracted from NN data using various chiral potentials 
vary with the cutoff in the LSE. The cutoff-dependence of these observables, as well as of the phase shifts at $T_{lab} \approx 100$ MeV, suggests that for a chiral potential computed
with dimensional regularization the highest LSE cutoff it is sensible to adopt is approximately 1 GeV. Using spectral-function regularization to compute the two-pion-exchange potentials postpones the onset of cutoff dependence in these quantities, but does not remove it. 
\end{abstract}

\maketitle

\vspace{10mm} 
\section{Introduction}

In the present work we will focus on nucleon-nucleon (NN)
scattering in chiral effective theory ($\chi$ET). This effective theory is based on the spontaneously broken chiral symmetry of QCD.
In it the long-range part of the strong interaction between nucleons is given by the exchange of QCD's approximately massless Goldstone bosons---the pions.
This dynamics is described by an effective Lagrangian which inherits the
$SU(2)_L \times SU(2)_R$ symmetry of QCD---as well as the pattern of its breaking---and so reproduces QCD in the energy region below the chiral-symmetry-breaking scale $\Lambda_{\chi {\rm SB}} \approx 1$ GeV. In order for the effective theory to accurately represent QCD
the low-energy constants that encode the effects of degrees of freedom with energies  $>\Lambda_{\chi {\rm SB}}$ must be determined from either 
lattice QCD calculations or experimental data. 

Since this theory is a quantum field theory, it requires renormalization. 
But, as an effective theory, it contains non-renormalizable operators. As a result one needs to 
establish a power-counting scheme in order to renormalize the theory order by order. 
The usual chiral counting in the single-nucleon sector involves a perturbative expansion 
in powers of $Q$ (``chiral perturbation theory") where $Q$ is either the momentum of the 
particles involved or the pion mass: $Q=\{p,m_\pi\}$. 
Weinberg \cite{We90,We91} showed that a similar $\chi$PT expansion could be made for the 
NN potential, although infra-red enhancements due to the presence of two heavy particles 
rendered such a perturbation theory questionable for the NN amplitude.
Since then considerable efforts have been devoted to the construction of NN potentials that 
are based on a $\chi$PT expansion~\cite{Or96, Ka97, Ep99, EM03, Ep05}, with calculations now 
complete up to N$^3$LO (=$O(Q^4)$). These potentials have both long-distance ($r \sim 1/m_\pi$) 
and short-distance ($r \ll 1/m_\pi$) parts. A key feature of the resulting ``chiral effective 
theory" is that only certain short-distance structures are permitted at a given order in the $Q$ 
expansion, because a momentum expansion is being employed. 
 
With an NN potential available it would seem straightforward to iterate that potential in a
Lippmann-Schwinger equation (LSE) and so obtain the NN scattering amplitude, and from there
 phase shifts~\cite{Or96, Le97, Ep99,Ge99,EM02,EM03,Ol03,Ep05,Dj07}. 
How to perform such a non-perturbative treatment consistently remains an open question.  It requires a cutoff 
in the LSE, because the $\chi$PT potentials grow with the exchange momentum ${\bf q}$. 
The theory cannot be regarded as \textquotedblleft properly
renormalized\textquotedblright\ unless it produces results that are 
cutoff independent---or contains only weak cutoff dependence that can be removed by including 
higher-order short-distance operators.
A contrast to such use of potentials in the LSE is provided by
attempts to establish a perturbative expansion for nuclear forces that incorporates chiral symmetry. 
But straightforward implementation of this perturbation theory results in poor convergence in the 
key ${}^3$S$_1$-${}^3$D$_1$ 
channel~\cite{Kaplan:1998tg, Kaplan:1998we, Fleming:1999ee, Fleming:1999bs} (c.f. the recent work of 
Ref.~\cite{Be08}). This failure of perturbation theory is associated with the appearance of a 
new scale in the NN problem---one associated with the tensor part of the NN 
force~\cite{Kaplan:1998we,Be02}. Renormalization-group methods can then be used to organize and 
implement a mixed treatment~\cite{BMcG04,BB03,Bi06}, where one-pion exchange is treated 
non-perturbatively, and all parts of the long-range force that are of higher chiral order are 
evaluated in distorted-wave Born approximation (see also Ref.~\cite{LvK08}). But, in any of these 
analyses where pion effects are treated non-perturbatively,  the complexity of the potential means 
that all analyses of the residual cutoff dependence of results are necessarily numerical. 

The operators of lowest chiral order that appear in the NN potential are two S-wave contact 
interactions, together with the one-pion-exchange potential. These operators are all $O(Q^0)$ and 
have been regarded as forming the ``leading-order" potential. In Refs.~\cite{Be02,PVRA04A,PVRA04B} 
it was shown the theory is properly renormalized in the $^{1}$S$_{0}$, $^{3}$S$_{1}-^{3}$D$_{1}$ 
channels and Ref.~\cite{NTvK05} found that the same was true in some of the higher partial 
waves ($^{1}$P$_{1}$, $^{1}$D$_{2}$, $^{1}$F$_{3}$, $^{1}$G$_{4}$, $^{3}$P$_{1}$, 
$^{3}$F$_{3}$)~\footnote{Here we do not discuss the issue of quark-mass dependence, although 
it should be noted that iteration of a potential of fixed $\chi$PT order also leads to 
inconsistencies as regards the counting of short-distance operators proportional to the 
quark-mass~\cite{KSW96,Be02}.}. 
In contrast, Ref.~\cite{NTvK05} showed that $\chi$ET is 
not properly renormalized at LO in spin-triplet channels where an attractive singular tensor 
force acts (e.g. $^{3}$P$_{0}$, $^{3}$D$_{2}$, $^{3}$P$_{2}-^{3}$F$_{2}$) 
(see also Ref.~\cite{ES01}). In Ref.~\cite{NTvK05} it was suggested that a
contact term needs to be added in these channels---even though this breaks the straightforward 
$\chi$PT counting for short-distance operators. This behavior in different NN partial waves can be 
understood as a result of the singular behavior of the $\chi$ET NN potential, with the channels 
where the leading singularity of the potential as $r \rightarrow 0$ corresponds to an attractive 
force needing a counter term to stabilize the phase-shift predictions as a function of the 
cutoff~\cite{Be01,NTvK05}. Partial waves where the leading singularity has a positive 
coefficient---and so the dominant effect at short distances is repulsion---have $\Lambda$-independent 
predictions at large $\Lambda$~\cite{PVRA06A,PVRA06B}. A renormalization-group analysis confirmed 
these findings~\cite{Bi06}. Ref.~\cite{EM06} questioned the usefulness of conclusions based 
on a leading-order analysis, as well as the necessity of considering cutoffs $> 1$ GeV.

In this paper we extend the analysis of the $\chi$ET potential in the NN P-waves at large cutoffs 
to the NLO and NNLO potentials derived in Refs.~\cite{Or96,Ka97,Ep99}. The key issue we are 
concerned with is whether renormalization of the LSE can be successfully accomplished when these 
higher-order potentials are employed. This problem has been addressed in Refs.~\cite{PVRA06B,En07} 
using NLO, NNLO, and N$^3$LO potentials and a co-ordinate space analysis. However, Ref.~\cite{En07} 
focused on the ${}^1$S$_0$ channel, and the analysis of Ref.~\cite{PVRA06B} only yields results 
for $\Lambda \rightarrow \infty$ and cannot address the approach to that limit. We know from the 
detailed studies of Refs.~\cite{Ep99,EM03,Ep05} that there exists a range of 
cutoffs $\Lambda=600$--$800$ MeV where the renormalization of the integral equation can be 
successfully carried out. However, if this analysis cannot be extended to cutoffs that are large with 
respect to the chiral-symmetry-breaking scale we are forced to conclude that use of the $\chi$PT 
potential as a non-perturbative kernel in the Lippmann-Schwinger equation only makes sense at 
cutoffs $\Lambda < \Lambda_{\chi {\rm SB}}$. For practical application of chiral potentials to 
problems in, e.g., nuclear structure or electromagnetic reactions, knowledge of 
the maximum sensible value of $\Lambda$ in such a treatment is a 
crucial component of forming accurate error estimates at a given order.

At next-to-leading (NLO) and
next-to-next-to-leading order (NNLO), the $\chi$PT potential acquires two-pion-exchange (TPE) parts, 
as well as seven short-distance operators of chiral dimension two, four of which are in 
one-to-one correspondence with contact interactions in the NN P waves~\cite{Ep99}. 
Here we will analyze the P-wave predictions due to these potentials using a method based on 
subtractive renormalization of the LSE~\cite{HM99,AP04,Ya08}. In this approach, we do not fit 
the constant that multiplies the short-distance operator in the NN potential directly to data. 
Instead, we perform manipulations on the LSE in order to obtain an integral equation where we can 
directly input a (generalized)
scattering length $\alpha_{11}$ which is obtained from low-energy P-wave phase
shifts. (An earlier, alternative subtractive-renormalization technique involves invoking the 
Born approximation, and consequently is unreliable for the highly-singular potentials being 
discussed here~\cite{Fr99,Ti05}.)
Using this method we can predict P-wave phase shifts at cutoffs up to $10$ GeV, and examine the 
cutoff dependence of these predictions. 
We then examine how far $\alpha_{11}$ must be varied in order to obtain a best fit to 
phase shifts below $T_{lab}=100$ MeV for each
cutoff. These numbers are compared with the $\alpha_{11}$'s obtained from the low-energy phase-shift
analysis, in order to see whether there is evidence for a critical value of the cutoff at which it is difficult to fit both the threshold data and the data in the range up to $100$ MeV. We also
perform the same calculation but replace the $Q^{3}$ contribution (and later, the whole) to TPE by
the spectral-function regularization (SFR)~\cite{epsfr} results with a SFR cutoff 
$\tilde{\Lambda}=800$~MeV therein. 

The structure of our work is as follows. First, in Section~\ref{sec-subtract}, we introduce 
our subtractive renormalization scheme for P-waves. Then, in Section~\ref{sec-ope}, we consider 
the results when one-pion exchange alone is taken as the long-range potential in these partial waves, 
and reproduce the results of Ref.~\cite{NTvK05} with our subtractive renormalization method. 
In Section~\ref{sec-tpe}, we list the TPE obtained via dimensional regularization (DR) and 
spectral-function regularization (SFR) and the relevant partial-wave
decomposition formulas. In Section~\ref{sec-result}, we use these potentials, together with the 
subtracted LSE, to obtain phase shifts for all four NN P waves over a range of cutoffs from 500~MeV 
to 10~GeV. We then discuss the implications of our results for $\chi$ET in the NN system and
 summarize our findings in Section~\ref{sec-con}.


\section{Subtractive renormalization in P-waves}
\label{sec-subtract}

In this section we will introduce an approach to the renormalization of the Lippmann-Schwinger 
equation that is based on subtractive renormalization~\cite{AP04,Ya08}. 
This has the advantage that only renormalized quantities appear in the final 
equation, with all reference to the coefficient of the contact interaction, the NN low-energy
constants (LECs), having 
disappeared.  Such techniques are in general useful for potentials that are the sum of a 
long-range part and a contact interaction. 
The difference between Ref.~\cite{Ya08} and this work is that, in contrast to the constant contact 
interactions considered there, we now treat contact interactions that are operative in P-waves, 
and thus are momentum dependent.

The partial-wave LSE has the following form
\begin{equation}
t_{l^{\prime }l}(p^{\prime },p;E)=v_{l^{\prime }l}(p^{\prime
},p)+\sum_{l^{\prime \prime }}\frac{2}{\pi }M\int_{0}^{\Lambda }
\frac{dp^{\prime \prime }\;p^{\prime \prime }{}^{2}\;v_{l^\prime l^{\prime \prime}}
(p^{\prime },p^{\prime \prime })\;t_{l^{\prime \prime }l}(p^{\prime \prime},p;E)}
{p_{0}^{2}+i\varepsilon -p^{\prime \prime}{}^{2}},  
\label{eq:7}
\end{equation}
where $p_{0}^{2}/M=E$ is the center-of-momentum (c.m.) energy and $\Lambda $ the cutoff parameter.
The angular momenta of the incoming (outgoing) momenta are indicated by the indices $l$ ($l'$).
We consider potentials of the form:
\begin{equation}
v_{l^{\prime }l} ^{SJ}(p',p) =v^{LR}_{l' l}(p',p) + C^{SJ}_{l' l} p'^{l'} p^l,
\label{eq:vlprimel}
\end{equation}
where $p (p^{\prime })$ indicates the incoming (outgoing) momentum in the 
center-of-momentum (c.m.) frame, and 
$v^{LR}_{l' l}$ is the long-range potential that is operative in this channel. 
For the purposes of this work $v^{LR}_{l' l}$ consists of at least one-pion exchange,
and, in most cases, includes two-pion exchange too. The second term is the simplest contact interaction 
that can appear after partial-wave decomposition in this channel.  We explicitly express the dependence
of this term on the quantum numbers $S$ and $J$, indicating that for each spin and total-angular-momentum channel there is,
in principle, a different constant. 

If operators up to $O(Q^3)$ are retained in the $\chi$ET potential then the four operators 
necessary to yield contact interactions in the NN
P-waves ($^{1}$P$_{1},$ $^{3}$P$_{0},$ $^{3}$P$_{1},$ $^{3}$P$_{2}-^{3}$F$_{2}$) are present---and indeed 
are needed in order to renormalize the $\chi$PT loop diagrams in $v^{LR}_{l' l}$ that 
diverge as ${\bf q}^{2}$. 
The piece of these interactions that survives after projection onto the $l=l'=1$ partial wave 
is an operator proportional to ${\bf p}' \cdot {\bf p}$,
which produces the second term in Eq.~(\ref{eq:vlprimel}) in that case.

The main idea of our subtraction method is to construct the fully off-shell partial-wave 
$t$-matrix from the knowledge of $v^{LR}$ and the on-shell value of the $t$-matrix
for zero energy.  This procedure of subtracting
LSEs at zero energy to cancel unknown constants was introduced in
Ref.~\cite{HM99}  for s-waves and
will be generalized in the following for partial waves with $l\neq 0$. 

First, a generalized scattering length for arbitrary angular momenta $l$ and $l^\prime$ can be
defined as
\begin{equation}
\frac{\alpha _{l^\prime l}^{SJ}}{M}=
\lim_{k\rightarrow 0}\frac{t_{l^{\prime}l}^{SJ}(k,k;E)}{k^{l^{\prime }+l}}.  
\label{eq:8}
\end{equation}
Dividing the partial-wave LSE, 
Eq.~(\ref{eq:7}), by $p'^{l'} p^l$ we obtain
\begin{equation}
\frac{t_{l^{\prime }l}^{SJ}(p^{\prime },p;E)}{p'^{l'} p^l}=
\frac{v_{l^{\prime}l}^{SJ}(p^{\prime },p)}{p'^{l'} p^l}+\sum_{l^{\prime \prime }}
\frac{2}{\pi }
\frac{M}{p'^{l'} p^l}\int_{0}^{\Lambda }
\frac{dp^{\prime \prime}\;p^{\prime \prime }{}^{2}\;
v_{l^\prime l^{\prime \prime }}^{SJ}(p^{\prime },p^{\prime
\prime })\;t_{l^{\prime \prime }l}^{SJ}(p^{\prime \prime },p;E)}
{p_{0}^{2}+i\varepsilon -p^{\prime \prime }{}^{2}}.  
\label{eq:8.5}
\end{equation}

\noindent
Since $v^{LR}_{l^{\prime }l}(p',p)\sim p'^{l^{\prime
}}p^{l}$, the ansatz of Eq.~(\ref{eq:8.5}) is general and can be applied to any partial wave. In the following we 
concentrate on P-waves ($l=l^\prime=1$). For ease of notation
we drop for the derivation the indices referring to the angular momenta $S$ and $J$, unless they are explicity 
needed.
In the case of P-waves the full potential $v_{l^{\prime}l}^{SJ}(p^\prime ,p)$ involves a contact interaction of the form 
$C_{l^\prime l}^{SJ} \; p^\prime p$. First we consider the half-shell and on-shell $t$-matrices at zero energy:

\begin{eqnarray}
\lim_{k\rightarrow 0} \left[ \frac{t_{l^\prime l}(p^{\prime },k;0)}{p^{\prime }k} \right]
&=&\lim_{k\rightarrow 0}\left[ \frac{v^{LR}_{l^\prime l }(p^{\prime },k)}{p^{\prime }k}
+C_{l^\prime l} \right] \cr
&+& \sum_{l^{\prime \prime }}\frac{2}{\pi }M\lim_{k\rightarrow 0} \left[ \frac{1}{%
p^{\prime }k}\int_{0}^{\Lambda }\frac{dp^{\prime \prime }\;p^{\prime \prime
}{}^{2}\;(v^{LR}_{l^\prime l^{\prime \prime} }(p^{\prime },p^{\prime \prime })+
C_{l^\prime l^{\prime \prime}}p^{\prime }p^{\prime
\prime })\;t_{l^{\prime \prime }l}(p^{\prime \prime },k;0)}{-p^{\prime
\prime }{}^{2}} \right] 
\label{eq:9} \\
\lim_{k\rightarrow 0}\left[ \frac{t_{l^{\prime }l}(k,k;0)}{kk} \right]
&=&\lim_{k\rightarrow 0}\left[ \frac{v^{LR}_{l^\prime l }(k,k)}{kk}+C_{l^\prime l}\right] \cr
&+&  \sum_{l^{\prime \prime
}}\frac{2}{\pi }M\lim_{k\rightarrow 0}\left[ \frac{1}{kk}\int_{0}^{\Lambda }\frac{%
dp^{\prime \prime }\;p^{\prime \prime }{}^{2}\;(v^{LR}_{l^\prime l^{\prime \prime} }(k,p^{\prime \prime
})+C_{l^\prime l^{\prime \prime}}kp^{\prime \prime })\;
t_{l^{\prime \prime }l}(p^{\prime \prime },k;0)}{-p^{\prime \prime }{}^{2}} \right].  
\label{eq:10}
\end{eqnarray}

\noindent
Subtracting Eq.~(\ref{eq:10}) from Eq.~(\ref{eq:9}) and multiplying both
sides by $p^{\prime }$, the unknown constant $C_{l^\prime l}$ cancels and we obtain
\begin{eqnarray}
\lim_{k\rightarrow 0} \left[ \frac{t_{l^{\prime }l}(p^{\prime },k;0)}{k} \right]
-\frac{\alpha _{11}}{M}p^{\prime } 
& =& \lim_{k\rightarrow 0} \left[ \frac{v^{LR}_{l^\prime l}(p^{\prime },k)}{k} \right]
-p^{\prime }\lim_{k\rightarrow 0} \left[ \frac{v_{l^\prime l}^{LR}(k,k)}{
k^{2}} \right] \nonumber \\
&+&\sum_{l^{\prime \prime }}\frac{2}{\pi }M\int_{0}^{\Lambda }\frac{
dp^{\prime \prime }\;p^{\prime \prime }{}^{2}\; \left[v^{LR}_{l^\prime l^{\prime \prime} }(p^{\prime
},p^{\prime \prime })-\lim_{k\rightarrow 0}[\frac{v^{LR}_{l^\prime l^{\prime \prime} }(k,p^{\prime
\prime })}{k} ] p^{\prime } \right]\lim_{k\rightarrow 0}[\frac{\;t_{l^{\prime \prime
}l}(p^{\prime \prime },k;0)}{k}]}{-p^{\prime \prime }{}^{2}}.  \label{eq:11}
\end{eqnarray}

Here we have used that for P-waves ($l{^\prime} =l=1$), 
$\lim_{k\rightarrow 0}\left[\frac{t_{11}(k,k;0)}{kk}\right]=
\frac{\alpha_{11}}{M}$. Note that for the coupled channels $^{3}$P$_{2}-^{3}$F$_{2}$, 
counter terms only exist for $(l,l^{\prime})=(1,1)$, i.e., the $^{3}$P$_{2}-^{3}$P$_{2}$
part, if one considers operators up to $O(Q^3)$. Thus, up to $O(Q^3)$, the treatment for 
this channel is exactly the same as for the uncoupled channels, since there is no new 
unknown constant. All of the above limits are well defined and can be
directly obtained from Eqs.~(\ref{eq:1})-(\ref{eq:4}) in Section~\ref{sec-tpe}. The only unknown left in
Eq.~(\ref{eq:11}) is $\lim_{k\rightarrow 0}\left[\frac{t_{l^{\prime }l}(p^{\prime
},k;0)}{k}\right]$, and one can solve the equation by standard techniques. 

With the half-shell $t$-matrix at zero energy, 
 $\lim_{k\rightarrow 0}\left[\frac{t_{l^{\prime }l}(p^{\prime },k;0)}{k}\right]$,
in hand,
we can write another set of two LSEs to obtain the fully off-shell t-matrix,
\begin{eqnarray}
\frac{t_{l^{\prime }l}(p,p^{\prime };0)}{pp^{\prime }} &=&\frac{v_{l^\prime
l}^{LR}(p,p^{\prime })}{pp^{\prime }}+C_{l^\prime l}+\sum_{l^{\prime \prime }}
\frac{2}{\pi }M \left[
\frac{1}{pp^{\prime }}\int_{0}^{\Lambda }\frac{dp^{\prime \prime
}\;p^{\prime \prime }{}^{2}\;(v^{LR}_{l^\prime l^{\prime \prime} }(p,p^{\prime \prime })
+C_{l^\prime l^{\prime \prime} }pp^{\prime
\prime })\;t_{l^{\prime \prime }l}(p^{\prime \prime },p^{\prime };0)}{%
-p^{\prime \prime }{}^{2}}\right] 
 \label{eq:12} \\
\lim_{k\rightarrow 0}\left[\frac{t_{l^{\prime }l}(k,p^{\prime };0)}{kp^{\prime }}%
\right] 
&=&\lim_{k\rightarrow 0}\left[\frac{v^{LR}_{l^\prime l }(k,p^{\prime })}{kp^{\prime }}
+C_{l^\prime l}\right] \cr
&+& \sum_{l^{\prime \prime }}\frac{2}{\pi }M\lim_{k\rightarrow 0}\left[\frac{1}{
kp^{\prime }}\int_{0}^{\Lambda }\frac{dp^{\prime \prime }\;p^{\prime \prime
}{}^{2}\;(v^{LR}_{l^\prime l^{\prime \prime} }(k,p^{\prime \prime })+
C_{l^\prime l^{\prime \prime} }kp^{\prime \prime})\;
t_{l^{\prime \prime }l}(p^{\prime \prime },p^{\prime };0)}{-p^{\prime\prime }{}^{2}}\right].
\label{eq:13}
\end{eqnarray}
Subtracting  Eq.~(\ref{eq:13}) from Eq.~(\ref{eq:12}) and multiplying both
sides by $p$, the constant $C_{l^\prime l}$ again cancels and we arrive at
\begin{eqnarray}
\frac{t_{l^{\prime }l}(p,p^{\prime };0)}{p^{\prime}}-p\lim_{k\rightarrow 0}\left[\frac{
t_{l^{\prime }l}(k,p^{\prime };0)}{kp^{\prime }}\right]
&= &\frac{v^{LR}_{l^\prime l }(p,p^{\prime })}{p^{\prime }}-p\lim_{k\rightarrow 0}\left[\frac{v^{LR}_{l^\prime l }(k,p^{\prime })}{kp^{\prime }}\right] \cr
&+&\sum_{l^{\prime \prime }}\frac{2}{\pi }
M\int_{0}^{\Lambda }\frac{dp^{\prime \prime }\;p^{\prime \prime
}{}^{2}\left[v^{LR}_{l^\prime l^{\prime \prime} }(p,p^{\prime \prime })-p\lim_{k\rightarrow 0}[\frac{v^{LR}_{l^\prime l^{\prime \prime} }(k,p^{\prime \prime })}{k}]\right] 
\frac{\;t_{l^{\prime \prime }l}(p^{\prime
\prime },p^{\prime };0)}{p^{\prime }}}{-p^{\prime \prime }{}^{2}}.
\label{eq:14}
\end{eqnarray}

In the case $l'=l$ we can then use the property of the $t$-matrix that  $t_{l'l}(k,p^{\prime };0)=t_{l'l}(p^{\prime },k;0)$ to obtain $\lim_{k\rightarrow 0}[\frac{t_{l^{\prime
}l}(k,p^{\prime };0)}{kp^{\prime }}]$ and  solve Eq.~(\ref{eq:14}) to find 
$\frac{t_{l^{\prime }l}(p,p^{\prime };0)}{p}$. Once we obtain 
$\frac{ t_{l^{\prime }l}(p,p^{\prime };0)}{p}$, or, equivalently,  
$t_{l^{\prime }l}(p,p^{\prime };0)$, we can proceed to calculate the on-shell $t$-matrix 
and the phase shifts using resolvent 
identities that connect the operator $t(E)$ to the operator
$t(0)$. Those details are laid out in Refs.~\cite{Fr99,Ya08}.


\section{One-pion-exchange potential in p-waves}
\label{sec-ope}

In the chiral effective theory the operators of lowest chiral dimension which can contribute to 
the NN potential are the operators of the one-pion-exchange potential (OPE). 
The leading-order long-range potential is thus $v^{LR}=v_{1 \pi}$, which we consider in this 
section. It has the well known form
\begin{equation}
v_{1 \pi}({\bf p'},{\bf p})=-\tau_1 \cdot \tau_2 \frac{g_{A}^{2}}{4 f_\pi^2} 
\frac{{\bf \sigma}_1 \cdot {\bf q} \, {\bf \sigma}_2 \cdot {\bf q}}{m_{\pi }^{2}+\mathbf{q}^{2}}, 
 \label{eq:2.1}
\end{equation}
where $\mathbf{q}=\mathbf{p}-\mathbf{p}'$. When  considering the OPE alone, 
we adopt the axial vector coupling constant $g_{A}=1.29$, and pion decay constant 
$f_{\pi }=92.4$~MeV. (By choosing this value for $g_A$ in the OPE we are taking into account 
$O(Q^2)$ corrections to it that shift the $\pi$NN coupling constant away from the 
Goldberger-Treiman result and give us the freedom to reproduce the empirical value.)
In addition, the nucleon mass $M=938.926$~MeV and pion mass $m_{\pi }=138.03$~MeV are
employed.
In the S-wave channels,  $^{1}$S$_{0}$ and $^{3}$S$_{1}-^{3}$D$_{1}$, the OPE can be properly 
renormalized by a constant contact interaction in each of the channels~\cite{Be02,PVRA04A,PVRA04B}.
These contact interactions represent the ignorance of contributions from high momenta, 
i.e. short-range physics. However, since the lowest order in which a contact interaction 
that is  a polynomial of $\mathbf{q}$ can survive after partial wave decomposition is 
proportional to $\mathbf{q}^{l+l^{\prime}}$, there is no contact interaction in 
leading order in the P-waves unless one adds a higher-derivative operator
 to the leading-order potential, i.e. an operator  that is $O(Q^{2})$ in the 
usual $\chi$PT counting in powers of $\mathbf{q}$. 

Therefore, in the Weinberg prescription for computation of the $\chi$ET potential, the 
OPE alone gives the LO phase shifts in P-waves. The $v_{l'l}$ in each partial 
wave can then be obtained using the projection formulae presented in the next section. 
The resulting phase shifts in the four NN P-waves are shown
in the left panels of Figs.~\ref{fig1}-\ref{fig2}.
Fig.~\ref{fig1} shows that the phase shift of $^{1}$P$_{1}$ and $^{3}$P$_{1}$ 
become independent 
of the choice of cutoff once $\Lambda >2000$ MeV. However,  for the  attractive 
 channels $^{3}$P$_{0}$ and $^{3}$P$_{2}$-$^{3}$F$_{2}$, the phase shifts in
Fig.~\ref{fig2} are cutoff dependent. 
This reproduces the finding of Ref.~\cite{NTvK05}, where the attractive nature of the 
tensor $1/r^3$ potential in these channels leads to strong cutoff dependence of 
the phase shifts.  

To stabilize these channels, a contact interaction must be included. Thus, we include contact 
interactions that survive in the P-waves after partial-wave decomposition. 
Once we do this,  the subtractive renormalization described in the previous section can 
be employed and the LSE be solved. 
For our calculations we first employ the generalized $l=1$ scattering lengths extracted
 in~\cite{vald} as the input value to our subtraction scheme. (See first two lines of Table~\ref{table-0}.)
As shown in the right panels of Figs.~\ref{fig1}-\ref{fig2},
these contact interactions absorb the cutoff dependence 
in the $^{3}$P$_{0}$ and $^{3}$P$_{2}$-$^{3}$F$_{2}$ channels. However, a strong cutoff 
dependence is now created in the $^{1}$P$_{1}$ and $^{3}$P$_{1}$ channels, and
produces a resonance-like structure in the phase shifts. Note that since we adopt the
generalized scattering lengths extracted in~\cite{vald} to perform the subtraction, the
results are not necessarily the best fit to the Nijmegen phase shift analysis as we will
discuss in more detail in Section~\ref{sec-result}.

With the calculations shown in Figs.~\ref{fig1} and \ref{fig2} we 
reproduce the finding of Ref.~\cite{NTvK05}: 
if cutoffs larger than $\approx 600$~MeV are to be considered then additional contact interactions 
must be included in the LO potential which 
act in the ${}^3$P$_0$ and ${}^3$P$_2$ channels. Adding those contact interactions in the 
other P-waves destabilizes the phase-shift prediction as a function of cutoff---as might 
have been expected given the fact that the  OPE potential is either non-singular (${}^1$P$_1$) or singular and repulsive
(${}^3$P$_1$) 
as $r \rightarrow 0$ in these waves~\cite{PVRA06A,PVRA06B}.


\section{Two-pion-exchange potential}
\label{sec-tpe}

The two-pion-exchange potential (TPE), $v_{2 \pi}$, appears first at next-to-leading order 
(NLO) in 
the chiral effective theory. Here we will consider two parts of $v_{2 \pi}$: 
the  NLO ($Q^{2}$) part and the next-to-next-to-leading order (NNLO) part, which is $O(Q^{3})$. 
In the NLO piece the divergent loop diagrams must be regulated in order to obtain the 
potential. This can be achieved by either
dimensional regularization~\cite{Or96, Ka97, Ep99} or spectral-function regularization~\cite{epsfr}.
After regularization, counter terms up to $O(Q^{2}$) are added to the TPE to 
absorb the divergences. 
We first consider the TPE obtained via dimensional regularization (DR), and adopt 
the formulation of \cite{Ep99}, where the axial-vector
coupling constant is $g_{A}=1.26$, and the pion decay constant $f_{\pi }=93$~MeV. 
The $\pi$N LECs that appear in the NNLO potential are given by 
$c_{1}=-0.81$, $c_{3}=-4.70$, and  $c_{4}=3.4$ (all in GeV$^{-1}$)~\cite{Ep99}.
We use  the TPE as derived in Ref.~\cite{Ka97}, which includes the 
$1/M$ corrections. This is the correct form in the power counting 
$M \sim 4 \pi f_\pi \approx \Lambda_{\chi {\rm SB}}$~\cite{EM03}, although 
a different counting is adopted in Refs.~\cite{Ep05,epsfr}~\footnote{It could be argued that in order to achieve NNLO accuracy in this power counting we should also consider $1/M^2$ corrections to OPE. Ref.~\cite{Friar}
points out that these corrections can be accounted for by adopting the potential of Ref.~\cite{Ka97} and modifying $v_{1 \pi}$ by an energy-dependent factor. Using the prescription of Ref.~\cite{Friar} to include these $1/M^2$ corrections to
$v_{1 \pi}$ in our calculation alters our phase shifts by only a small amount at the energies under consideration here. It has also been shown in Ref.~\cite{spm} that these $1/M^2$ pieces of $v_{1 \pi}$ have only a small  
impact in the $^1$S$_0$ channel. A precise description of phase-shift data is not our goal here, so we do not consider this correction further in what follows.}.

For the convenience of the reader
the explicit expressions for the TPE~\cite{Ka97} are given below:
\begin{eqnarray}
v_{2\pi } &=&V_{C}+ \boldsymbol\tau _1 \cdot \boldsymbol\tau _2 \; W_{C}+[V_{S}+
\boldsymbol\tau _{1}\cdot \boldsymbol\tau_{2} 
\; W_{S}] \; {\boldsymbol\sigma} _1 \cdot \boldsymbol\sigma _2 \cr
&+&[V_{T}+\boldsymbol\tau_{1}\cdot \boldsymbol\tau_{2} \; W_{T}]\; 
\boldsymbol\sigma_1 \cdot \mathbf{q} \; \boldsymbol{\sigma }_{2}\cdot \mathbf{q}   \cr
&+& [V_{SO}+\boldsymbol\tau_{1}\cdot \boldsymbol\tau_{2} \; W_{SO}] \; 
i(\boldsymbol\sigma_1+\boldsymbol{\sigma }_{2})\cdot (\mathbf{q}\times \mathbf{p}) \cr
&+& [V_{Q}+\boldsymbol\tau _{1}\cdot \boldsymbol\tau_{2} \; W_{Q}]\; 
\boldsymbol\sigma_{1}\cdot (\mathbf{q}\times
\mathbf{p})\; \boldsymbol\sigma_{2}\cdot (\mathbf{q}\times \mathbf{p}).
\label{eq:1}
\end{eqnarray}
Here $\boldsymbol\sigma_{1} (\boldsymbol\tau_1)$ are the spin (isospin) matrices. 
The subscripts $C$, $S$, $T$, $SO$, and $Q$ 
indicate the Central, Spin-Spin, Tensor, Spin-Orbit and Quadratic spin-orbit components of 
the potential. The parts that cannot be absorbed into contact interactions have the explicit 
form as listed in Ref.~\cite{Ka97}. For the NLO $v_{2 \pi}$ we have for the long-range (constant and $q^2$ terms excluded)
part of the potential
{\begin{eqnarray}
W_{C} &=&
\left[4m_{\pi }^{2}(5g_{A}^{4}-4g_{A}^{2}-1)+q^{2}(23g_{A}^{4}-10g_{A}^{2}-1)+
\frac{48g_{A}^{4}m_{\pi }^{4}}{4m_{\pi }^{2}+q^{2}}\right]L(q)
\label{eq:3.1}
\end{eqnarray}
and
\begin{eqnarray}
V_{T} &=&-\frac{1}{q^{2}}V_{S}=\frac{3g_{A}^{4}}{64\pi ^{2}f_{\pi }^{4}}
L(q),
\label{eq:3}
\end{eqnarray}
where 
\begin{equation}
L(q) =\frac{w}{q}\ln \frac{w+q}{2m_{\pi }},
\end{equation}
and  $w=\sqrt{4m_{\pi}^{2}+q^{2}}$ and $q=|{\bf q}|$.  
Meanwhile, the $O(Q^3)$ part of $v_{2 \pi}$ has contributions
\begin{eqnarray}
V_{C} &=&\frac{3g_{A}^{2}}{16\pi f_{\pi }^{4}} \bigg\{ 4(c_{1}-c_{3})m_{\pi }^{3}-
\frac{g_{A}^{2}m_{\pi }}{16M}(m_{\pi }^{2}+3q^{2})-\frac{g_{A}^{2}m_{\pi
}^{5}}{16M(4m_{\pi }^{2}+q^{2})}-c_{3}m_{\pi }q^{2} \cr
& & +\left[2m_{\pi }^{2}(2c_{1}-c_{3})-q^{2}(c_{3}+\frac{3g_{A}^{2}}{16M})\right]
(2m_{\pi}^2+q^2)A(q) \bigg\},
 \label{eq:3.2}
\end{eqnarray}
where
\begin{equation}
A(q) = \frac{1}{2q}\arctan \frac{q}{2m_{\pi }},\label{eq:3.3}
\end{equation}
and
\begin{eqnarray}
W_{C} &=&\frac{g_{A}^{2}}{128\pi Mf_{\pi }^{4}}\bigg\{(8-11g_{A}^{2})m_{\pi
}^{3}+(2-3g_{A}^{2})m_{\pi }q^{2}-\frac{3g_{A}^{2}m_{\pi }^{5}}{4m_{\pi
}^{2}+q^{2}} \cr
& & +\left[4m_{\pi }^{2}+2q^{2}-g_{A}^{2}(4m_{\pi }^{2}+3q^{2})\right](2m_{\pi
}^{2}+q^{2})A(q)\bigg\} \label{eq:3.4}
\end{eqnarray}
\begin{equation}
V_{T}=-\frac{1}{q^{2}}V_{S}=-\frac{9g_{A}^{4}}{512\pi Mf_{\pi }^{4}}\big\{ m_{\pi
}+(2m_{\pi }^{2}+q^{2})A(q)\big\}
\end{equation}
\begin{equation}
W_{T}=-\frac{1}{q^{2}}W_{S}=\frac{g_{A}^{2}}{32\pi f_{\pi }^{4}}\bigg\{(c_{4}+
\frac{2-3g_{A}^{2}}{8M})m_{\pi }+\left[(c_{4}+\frac{1}{4M})(4m_{\pi }^{2}+q^{2})-
\frac{g_{A}^{2}}{8M}(10m_{\pi }^{2}+3q^{2})\right]A(q)\bigg\}
\end{equation}
\begin{equation}
V_{SO}=\frac{3g_{A}^{4}}{64\pi Mf_{\pi }^{4}}\{m_{\pi }+(2m_{\pi
}^{2}+q^{2})A(q)\}
\end{equation}
\begin{equation}
W_{SO}=\frac{g_{A}^{2}(1-g_{A}^{2})}{64\pi Mf_{\pi }^{4}}\{m_{\pi }+(4m_{\pi
}^{2}+q^{2})A(q)\}  \label{eq:4}
\end{equation}

\noindent
Note that although the NNLO potential behaves as $|{\bf q}|^{3}$ at large $|{\bf q}|$, 
the counter terms that renormalize the TPE are only of $O(Q^2)$. 

Another way to regularize the infinite loop during the calculation of the TPE is to adopt 
a spectral-function regularization (SFR), which is designed to keep the contribution from 
the loop integral to energy-momenta that are in the region where $\chi$PT is manifestly applicable. 
The spectral-function regularized TPE has a form similar to that given above for the
TPE obtained with DR, but the functions $L(q)$ and $A(q)$ in 
Eqs.~(\ref{eq:3.1})--(\ref{eq:4}) are replaced by $L^{\widetilde{\Lambda}
}(q)$ and $A^{\widetilde{\Lambda}}(q)$, which have the following form~\cite{epsfr}:
\begin{eqnarray}
L^{\widetilde{\Lambda}}(q) &=&\theta (\widetilde{\Lambda} -2m_{\pi })\frac{w}{2q}\ln \left[\frac{\widetilde{\Lambda}
^{2}w^{2}+q^{2}s^{2}+2\widetilde{\Lambda} qws}{4m_{\pi }^{2}(\widetilde{\Lambda} ^{2}+q^{2})}\right],
\label{eq:s1} \\
A^{\widetilde{\Lambda}}(q) &=&\theta (\widetilde{\Lambda} -2m_{\pi })\frac{1}{2q}\arctan \left[\frac{%
q(\widetilde{\Lambda} -2m_{\pi })}{q^{2}+2\widetilde{\Lambda} m_{\pi }}\right],  \label{eq:s2}
\end{eqnarray}
where $\widetilde{\Lambda}$ denotes the SFR cutoff on the loop integrals that arise when evaluating the potential and $s=\sqrt{\widetilde{\Lambda} ^{2}-4m_{\pi }^{2}}$.

By defining the linear combinations
 $U_{K}=V_{K}+(4I-3)W_{K}$ ($K=C,S,T,SO,Q$), with $I$ being the
 total isospin, the partial-wave form of the potential 
can be written as~\cite{Ep99}:
\begin{equation}
v_{l^{\prime }l}(p^{\prime },p)=\langle l^{\prime}sj|v(p^{\prime },p)|lsj \rangle,  \label{eq:5}
\end{equation}
where $l,l^{\prime } (s)$ denote the angular-momentum (spin) quantum numbers. 
With this we list the different partial-wave potentials for the TPE~\cite{Ep99}

\begin{eqnarray}
\langle j0j|v(p^{\prime },p)|j0j \rangle=\frac{-1}{8\pi }\int_{-1}^{1}dz
\left\{U_{C}-3U_{S}+p^{\prime }p(z^{2}-1)U_{Q}-q^{2}U_{T}\right\}P_{j}(z),
\label{eq:5.1}
\end{eqnarray}
\begin{eqnarray}
\langle j1j|v(p^{\prime },p)|j1j \rangle=\frac{-1}{8\pi }\int_{-1}^{1}dz
\bigg\{ \left[U_{C}+U_{S}-4p^{\prime }pzU_{SO}-p^{\prime
2}p^{2}(1+3z^{2})U_{Q}+(p^{\prime 2}+2p^{\prime }pz+p^{2})U_{T}\right] P_{j}(z) \nonumber \\
+\left[2p^{\prime }pU_{SO}+2p^{\prime 2}p^{2}zU_{Q}-2p^{\prime
}pU_{T}\right] (P_{j-1}(z)+P_{j+1}(z))\bigg\} 
\label{eq:5.2}
\end{eqnarray}
\begin{eqnarray}
\langle j\pm 1,1j|v(p^{\prime },p)|j\pm 1,1j \rangle=\frac{-1}{8\pi }\int_{-1}^{1}dz
\bigg\{p^{\prime }p \left[ 2U_{SO}\pm \frac{2}{2j+1}(-p^{\prime
}pzU_{Q}+U_{T})\right] P_{j}(z) \cr
+\left[ U_{C}+U_{S}-2p^{\prime }pzU_{SO}+p^{\prime
2}p^{2}(1-z^{2})U_{Q} \pm \frac{1}{2j+1}(2p^{\prime 2}p^{2}U_{Q}-(p^{\prime
2}+p^{2})U_{T})\right] P_{j\pm 1}(z)\bigg\} 
\label{eq:5.3}
\end{eqnarray}
\begin{eqnarray}
\langle j\pm 1,1j|v(p^{\prime },p)|j\mp 1,1j \rangle=\frac{-1}{8\pi }\frac{\sqrt{j(j+1)}
}{2j+1}\int_{-1}^{1}dz\bigg\{-4p^{\prime }pU_{T}P_{j}(z)+\left[\mp \frac{2p^{\prime
2}p^{2}}{2j+1}U_{Q}+2p^{\prime 2}U_{T}\right] P_{j\mp 1}(z) \cr
+\left[\pm \frac{2p^{\prime 2}p^{2}}{2j+1}U_{Q}+2p^{2}U_{T}\right] P_{j\pm 1}(z)\bigg\}.
\label{eq:6} 
\end{eqnarray}


\section{Results with different two-pion-exchange potentials}
\label{sec-result}

In this Section we present the P-wave phase shifts obtained with the TPE of $\chi$ET. 
They are obtained by solving the LSE via the subtractive renormalization
introduced in Section~\ref{sec-subtract}. We organize our findings as follows: 
we will discuss three different versions of the TPE, each of which is supplemented by the 
OPE that was the focus of Section~\ref{sec-ope}. In Subsections A and B we consider the TPE obtained via dimensional 
regularization up to NLO and NNLO respectively. In Subsection C, we consider the NNLO TPE obtained 
by spectral-function regularization. 
For each of these three potentials we first consider the ``bare" TPE, i.e. we solve the 
LSE with $v=v_{1 \pi} + v_{2 \pi}$ alone to  see how strong the cutoff dependence of the phase 
shift is for this ``un-renormalized" case. Then, we perform our subtractive renormalization, 
thereby incorporating contact interactions $C^{SJ}_{l l'} p^l p'^{l'}$ into the potential. 
We first solve the problem by adopting the generalized scattering lengths $\alpha^{SJ} _{11}$ 
extracted in Ref.~\cite{vald}. Then we adjust $\alpha^{SJ} _{11}$ to find the best fit with 
respect to the Nijmegen phase-shift analysis~\cite{St93}. Throughout
this work we only consider $np$ phase shifts, since we are mainly concerned
with issues of renormalization of the NN potential obtained from 
$\chi$ET. While the subtleties of isospin breaking can be computed
systematically within this framework, they go beyond the scope of this
work.

The issue of whether a contact term is required for a singular potential to reach cutoff independence can be linked to the short distance behavior of the potential in the coordinate space (see, for example, Refs.~\cite{Be01,PVRA06A,PVRA06B}). If the potential is singular and attractive at $r \rightarrow 0$, then the contact term is required, and if it is not then the phase shifts will have a stable $\Lambda \rightarrow \infty$ limit even in the absence of a contact term. In Subsection D, we summarize our findings for the different chiral potentials and link them to the potentials' coordinate-space behavior.

\subsection{Dimensionally regularized TPE at NLO}
\label{sec-a}

According to the power counting in powers of $Q$, the TPE at NLO in P-waves should
 be associated with  a contact interaction $C_{11}^{SJ}pp^{\prime}$. To obtain insight into
its effect, we first adopt the ``bare" TPE at NLO---i.e. we do not consider the contact
interactions when solving the LS equation---and  examine the $\Lambda$ dependence of 
the resulting phase-shift predictions. For this case, the ``bare" TPE contains the constant and $q^2$ terms excluded in Eqs.~(\ref{eq:3.1}) $\&$ (\ref{eq:3}). The importance of these terms diminishes as $\Lambda$ is increased, with the effects becoming negligible once $\Lambda > 3$ GeV.

The results are shown as a function of energy in Fig. \ref{fig3}. Plotting the phase shift in each channel as a function of the cutoff (see Fig.~\ref{figunrealla}, black dotted line) makes it clear that the  
$^{3}$P$_{2}$ and $^{3}$P$_{1}$ partial waves show strong cutoff dependence, whereas 
the $^{1}$P$_{1}$ and $^{3}$P$_{0}$ phase shifts converge for cutoffs $\Lambda >1200$~MeV. 
This feature is important because once a potential reaches cutoff 
independence in the LS equation
adding further contact interactions may destroy this feature.

Now, in order to explicitly see the effect of the counter terms $C_{11}^{SJ}pp^{\prime}$, we adopt 
our subtractive method to solve the LSE. 
For this we need as input values the four generalized scattering lengths: 
$\lim_{k\rightarrow 0}[\frac{t_{1 1}(k,k;0)}{kk} ] \equiv \frac{\alpha _{11}}{M}$. 
For these we first adopt numerical values  that are between those of the sets given in the first two lines of Table~\ref{table-0}.  
That table lists the values obtained for the generalized P-wave scattering lengths 
from the NijmII and Reid93 potentials~\cite{St94}, together with those for the
Cd-Bonn~\cite{CDBONN} and AV18~\cite{AV18} potentials.

The results are shown in  Fig.~\ref{fig4}, and indicate that the phase shifts have less cutoff dependence than in the unrenormalized case---especially in  
the $T_{lab}<50$ MeV region. But it is not until $\Lambda > 5000$ MeV that most of the channels become cutoff independent, and the phase shift in the ${}^3$P$_2$ partial wave does not even stabilize then. In particular, in the $^1$P$_1$ and $^3$P$_0$ channels the contact interaction delays until $\Lambda \approx 2000$ MeV the convergence with respect to the cutoff that was obtained in its absence. Moreover, Fig.~\ref{fig4} makes clear that using the constants $\alpha_{11}^{SJ}$ 
extracted from the potentials of Ref.~\cite{St94} together with the NLO $\chi$ET potential does not yield a good description of the 
energy dependence of the Nijmegen phase shifts---even for laboratory energies  $T_{lab} < 100$~MeV. 

In order to obtain a better description, we vary $\alpha _{11}^{SJ}$ at each cutoff so as to match the Nijmegen 
phase shifts for $T_{lab}<100$~MeV. This leads to  the phase shifts shown in Fig.~\ref{fig5}. In general, after performing  our best fit we are able to match the Nijmegen P-wave  phase shifts much better, especially for $T_{lab}<60$~MeV. As we will show in the next subsection, the NNLO pieces of TPE are needed in order to improve the energy dependence of the predictions, and thereby allow a better fit at $T_{lab}>60$ MeV.

We now assess how our results change depending on whether $\alpha_{11}^{SJ}$ is input from Table~\ref{table-0} or if it is adjusted to fit the Nijmegen phase shifts in the range $T_{lab}=0$--$100$ MeV. If the results are significantly different then that means our effective theory has sizeable dependence on the choice of input---the renormalization-point dependence is large. 
In order to study this issue we denote the generalized scattering length which gives the
best description of a phase shift as $\alpha _{best}$. We then plot $\alpha_{best}$ against $\Lambda$ in
Fig.~\ref{figalla} and \ref{figalpha3p2}. For the $^3P_2$ case, the varation of $\alpha _{best}$ with $\Lambda$ is so large that it must be plotted separately. The huge deviation of $\alpha_{best}(^3$P$_2)$ from the NijmII value of $-0.28~{\rm fm}^3$ around $\Lambda=2500$ MeV is correlated with the lack of cutoff independence in that channel. The $\chi$ET fails at this order when it is pushed to such cutoffs. 

But problems occur already at $\Lambda \approx 1000$ MeV. 
This issue of a difference between the $\alpha_{best}$ necessary to fit phase shifts with the 
NLO TPE potential and the $\alpha_{11}^{SJ}$ obtained from ``high-precision" potentials 
is particularly obvious in the $^1$P$_1$ and $^3$P$_0$ channels, where at $\Lambda \approx 1400$ MeV and
$\Lambda \approx 1000$ MeV respectively, the phase-shift predictions diverge away from the Nijmegen phase shifts. 
At these particular cutoffs changes in the generalized scattering length cannot bring the $\chi$ET phase shifts any closer to the ``data": the use of any 
finite $\alpha _{11}^{SJ}$ as input results in almost the same phase shifts as a function of energy. 
(We have verified that the same results are obtained if a contact interaction $C^{SJ}_{11} p'p$ 
is added to the potential and the coefficient is fitted to data.) 

This feature is correlated with bumps that are seen when the NLO results for the best-fit ${}^1$P$_1$ and ${}^3$P$_0$ phase shifts at $T_{lab}=100$ MeV are  plotted versus cutoff (see black dotted line in the top two panels of Fig.~\ref{figallbest100}). 
Meanwhile, the lack of cutoff independence in the coupled ${}^3$P$_2$-${}^3$F$_2$ channel is prominent in the fourth, fifth, and sixth panel of Fig.~\ref{figallbest100}. The only wave here that has a stable phase shift at NLO after one subtraction with $\alpha_{11}$ adjusted to reproduce the low-energy ``data" is the ${}^3$P$_1$. Over a cutoff range of $\Lambda=600$--$3100$ MeV its 100 MeV phase shift varies by only a few degrees. 

The above results suggest that the highest LSE cutoff one can adopt if one inserts the NLO TPE in the 
integral equation is $1000$ MeV. As $\Lambda$ goes beyond this value we encounter increasing difficulties in reproducing the Nijmegen P-wave phase shifts.

\subsection{The NNLO TPE computed with dimensional regularization}
\label{sec-b}
We now adopt the ``bare" DR TPE at NNLO and examine the resulting cutoff dependence of 
the P-wave phase shifts. The results are shown in Fig.~\ref{figunrealla} and  Fig.~\ref{fig6}: all of the P-waves 
exhibit strongly cutoff-dependent features when $v^{LR}=v_{1 \pi} + v_{2 \pi}$ with $v_{2 \pi}$ 
computed up to $O(Q^3)$ using DR. Comparing this to the results obtained for the 
un-renormalized TPE at NLO in the previous subsection, we see that the NNLO part of TPE has 
caused the $^{1}$P$_{1}$ and $^{3}$P$_{0}$ phase shifts to oscillate with respect to the cutoff 
too.

Next, we add the contact interactions $C_{11}^{SJ} pp^{\prime}$ to the potential. We again first 
adopt values of $\alpha _{11}^{SJ}$ based on the first two rows of Table~\ref{table-0}. 
With this, the oscillation in the P-waves with respect to $\Lambda$ is greatly reduced for $\Lambda >2000$~MeV. 
However, 
as shown in Fig. \ref{fig7},
most of the resulting P-wave phase shifts are rather far away from the Nijmegen analysis.

To remedy the situation, we again vary the four input values $\alpha _{11}^{SJ}$ to find the 
best fit for $T_{lab}<100$~MeV and obtain the results shown in Fig. \ref{fig8}. At the cutoffs presented in Fig. \ref{fig8}, our best-fit
results are generally comparable in quality to those presented in Refs.~\cite{Ep99, EM02}. (Note that in \cite{EM02},
an exponential regulator is added to TPE, which further suppresses the TPE
at momentum $q\approx \Lambda $.) Results for 
$\alpha_{best}$ versus cutoff and the phase shifts at $T_{lab}=100$ MeV  for $\Lambda=600-3100$ MeV 
 are again plotted in Fig.~\ref{figalla} and Fig.~\ref{figallbest100}, respectively (red dashed line). For the $^1P_1$, $^3P_0$ and $^3P_1$ channels, there are only minor variations of the phase shift with respect to $\Lambda$. Those variations correspond exactly to the variation of $\alpha_{best}$ with $\Lambda$ shown in Fig.~\ref{figalla}. On the other hand, the $^3P_2-^3F_2$ phase shifts display oscillations with respect to $\Lambda$. At NNLO one subtraction is not enough to stabilize this channel.

Figure \ref{figalla} shows that changes of 20-40\% from the values of $\alpha _{11}^{SJ}$ obtained from the 
`high-precision' potentials are needed  in order to achieve the best fit. 
In particular,
$\alpha _{best}$\ shows oscillatory behavior
for $\Lambda =800-3100$ MeV in the $^{3}$P$_{0}$, $^{3}$P$_{1}$ and $^{3}$P$_{2}-^{3}$F$_{2}$ 
channels. The oscillations are $\approx$ 10\%, and appear to start or become larger once the cutoff increases above 1 GeV. 
Such changes in observable quantities with the cutoff are
disturbing, but are not beyond the bounds of what one might expect due to N$^3$LO corrections.

Now we examine the success of our fit of phase shifts  as $\Lambda$ changes.
Fig.~\ref{fig8} shows that
 there is no significant
difference in the phase shifts for those chosen cutoffs within the region where the fit was 
performed, namely $T_{lab}<100$~MeV.  
However, for $T_{lab}>100$ MeV, the phase shifts computed with a cutoff $\Lambda =500$~MeV  differ
from the ones with all other cutoffs for all four NN P-waves. This indicates that the 
cutoff of $500$~MeV  is too low: it cuts out too much of the TPE that enters the LSE. For 
$\Lambda =800-2000$~MeV, the phase shifts show little cutoff dependence apart from one
specific case: the $^{3}$P$_{2}-^{3}$F$_{2}$ in the immediate vicinity of $\Lambda =1200$ MeV. 
Here we are too close to the cutoff where these phase shifts diverge (see Fig.~\ref{figallbest100}).


\subsection{Spectral-function regularization TPE}
\label{sec-c}

Both the phase shifts and $\alpha_{best}$  should be (approximately) invariant over 
a range of cutoffs if we are to claim that the renormalization is done correctly.
The oscillation of $\alpha _{best}$ for $\Lambda > \Lambda _{\chi }$ makes us re-think the 
role played by the NNLO($Q^3$) part of the TPE in DR. This piece of the TPE contains 
terms proportional to $|{\bf q}|^{3}$,  which are non-polynomial and have no 
corresponding counter term. Although the results in Fig. \ref{fig8} show that the 
cutoff dependence of  the phase shifts was absorbed by the $O(Q^2)$ contact interaction, 
we still see that $\alpha_{best}$ depends rather strongly on $\Lambda$ in the $^3$P$_0$ 
and $^3$P$_1$ partial waves. Could  it be possible that the reason for this is the existence of 
a piece of the potential  that grows as $|{\bf q}|^3$, and that does not have a contact 
interaction to counter balance it?

If we employ spectral-function regularization (SFR) to compute TPE and fix the cutoff on the spectral function, $\widetilde{\Lambda}$,
to a particular value,
then the dominant power in the SFR TPE for $|{\bf q}| \gg \widetilde{\Lambda}$ is $|{\bf q}|^2$. The long-range potential thus now has the same large-$|{\bf q}|$ as the contact 
interaction.
Therefore,  we will investigate how the phase shifts obtained when replacing the $Q^{3}$ part of DR-TPE by 
SFR~\cite{epsfr} (with $\widetilde{\Lambda}=800$ MeV)  differ from those obtained in the previous Subsection. 
We focus particularly on whether a better convergence with respect to the LSE cutoff can be achieved. 
We call the resulting long-range potential in this case the ``mixed" NNLO TPE.

To see the effect of this ``mixed" potential, 
in Fig.~\ref{fig11}  we plot the P-wave phase shifts obtained by adjusting $\alpha _{11}^{SJ}$ to fit the phase shifts to the Nijmegen analysis for 
$T_{lab} < 100$ MeV. 
The most striking feature of the results is that a
pattern similar to that observed in the DR TPE at NLO appears in the $^{1}$P$_{1}$ 
and $^3$P$_{0}$ 
channels, i.e., there is no way to describe the Nijmegen phase shifts for cutoffs near
$\Lambda$~=~1400~(1000)~MeV for $^{1}$P$_{1} (^{3}$P$_{0})$. However, this is not very 
surprising, 
since the leading singularity of these potentials is now given by the TPE in NLO, and thus
 the large-cutoff behavior is the same as for that potential. 

We do find that the  oscillation in $\alpha_{best}$ with $\Lambda$ that is present for the DR NNLO TPE 
does not appear when the ``mixed" NNLO TPE is used:
the variations of $\alpha _{best}$ for all
P-waves are under $5\%$. In addition, the values of $\alpha _{best}$ are
closer to the values obtained from the `high-precision' potentials 
than are those found in the previous subsection.
But the attempt to tame the $Q^3$ part of the TPE at NNLO($Q^3$) by adopting the 
SFR NNLO pieces of the potential still fails for $\Lambda > 1000$ MeV, due to the issues in the 
$^{1}$P$_{1}$ and $^{3}$P$_{0}$ channels discussed in the previous paragraph.


Finally, we adopt the full SFR TPE up to NNLO with the inner cutoff $\widetilde{\Lambda}=800$ MeV, and present the best-fit results in Fig.~\ref{figallbest100} (phase shifts at 100 MeV vs. $\Lambda$) as well as in Fig.~\ref{figfullbest} (phase shifts vs. energy at different cutoffs). The use of SFR removes the oscillation behavior in the phase shifts that occurs with the NNLO DR TPE. The phase shifts at $T_{lab}=100$ MeV are now almost independent of cutoff from $\Lambda=600$ to $2500$ MeV. For $\Lambda>2500$ MeV the ``best-fitted" $^3P_0$ phase shift at $T_{lab}=100$ MeV becomes cutoff dependent. Thus, we conclude that for P-waves, once the SFR TPE is adopted up to NNLO, it is possible to perform the renormalization in the LS equation with a cutoff up to $2500$ MeV.

\subsection{The co-ordinate-space connection}

The results of the previous three subsections are summarized in Table~\ref{table-summary}.
In that table an entry ``R" implies that a contact interaction is needed to make
predictions in this channel independent of the cutoff. In contrast, an entry ``U" implies
that phase-shift predictions for that particular P-wave are already $\Lambda$ independent
before any contact interaction is added. (See Fig.~\ref{figunrealla} for a graphical representation of this distinction.) If, in spite of this, a contact interaction is
then added  for $\Lambda \gg \Lambda_{\chi {\rm SB}}$ the phase-shifts either 
develop a resonant structure or become insensitive to the input value of $\alpha_{11}^{SJ}$ used to fit the coefficient of the contact interaction. The entries ``U" correspond exactly to those cases where the leading $r \rightarrow 0$ singularity of the potential has a positive coefficient, and so a repulsive singular potential dominates the physics at large cutoffs. Conversely, entries ``R" occur in those channels where the leading $r \rightarrow 0$ singularity of the co-ordinate-space potential corresponds to attraction~\cite{PVRA06A, PVRA06B}. The ${}^3$P$_2-{}^3$F$_2$ channel for DR NNLO is a special case, since the coupled-channels potential has two attractive singular eigenpotentials in the $r \rightarrow 0$ limit. Hence one subtraction is not sufficient to make phase shifts independent of $\Lambda$ in this channel. The large-$\Lambda$ behavior of our calculation can thus be completely understood from this co-ordinate-space point of view.



\section{Summary and Conclusions}
\label{sec-con}

We have developed a subtractive renormalization scheme for $\chi$ET NN potentials with contact 
interactions of the form $C^{SJ}_{11} pp^{\prime}$ and carried out calculations of the NN P-waves 
with one and two pion exchanges. We employed four different types of TPE potential:
the dimensional-regularization (DR) TPE up to NLO and NNLO, the ``mixed" potential (DR TPE up to NLO 
plus NNLO pieces based on spectral-function regularization), and spectral-function regularization (SFR) up to NNLO. Calculations using these $\chi$ET NN potentials are done with and without the contact interaction. When the contact interaction is included our 
renormalization scheme only
requires the knowledge of the general scattering lengths $\alpha_{l^{\prime }l}^{SJ}$, and thus 
allows us to go to an arbitrarily high cutoff in the LS equation. 

In the OPE case, we reproduced the results of Ref.~\cite{NTvK05} with our subtractive renormalization
scheme, i.e., for the attractive triplet channels ($^{3}$P$_{0}$ and $^{3}$P$_{2}$-$^{3}$F$_{2}$) one 
needs to include a contact interaction that is nominally of higher chiral order so as to make the resulting phase 
shifts cutoff independent. We note that if such a counter term is added in {\it all} P-waves it destroys the 
cutoff independence in the $^{1}P_{1}$ and $^{3}P_{1}$ channels that is achieved with OPE alone. 

In case of the NLO (NNLO) TPE computed using DR we found that the generalized scattering lengths obtained from the NijmII
and  Reid93 
potentials in Ref.~\cite{vald} produce phase shifts that are (approximately) cutoff independent after $\Lambda >5000(2000)$ MeV. The only exception is the $^3$P$_2$-$^3$F$_2$ channel, where cutoff dependence persists to arbitrarily high cutoffs in both the NLO and NNLO cases. In the other channels, the final cutoff-independent phase shifts are rather far from those of the Nijmegen 
analysis---especially at NLO. We therefore vary the generalized scattering to obtain the best fit for phase shift under 100 MeV. After doing that, we found that for the NLO TPE cutoffs above $\Lambda$~=~1400(1000)~MeV need to be excluded in the $^1P_1$($^3P_0$) channel.  At these cutoffs the NLO $^{1}$P$_{1}$ and $^{3}$P$_{0}$ phase shifts
become insensitive to the input value of the constants $\alpha_{11}^{SJ}$. This occurs because the dominant piece of the co-ordinate-space potential as $r \rightarrow 0$ in these channels is repulsive.
Results at NLO also depend strongly on the renormalization point, since the energy dependence predicted there is rather unlike that of the data. 

At NNLO we are almost always able to fit all P-waves to the Nijmegen
analysis---at least at the level of accuracy obtained in the NNLO fit of Ref.~\cite{Ep99}---after adjusting 
the values of the constants $\alpha_{11}^{SJ}$. In order to achieve such a fit we must take $\Lambda \geq 600$ MeV, as otherwise the LSE cutoff removes too much of the TPE physics that drives the energy dependence of the phase shifts. But there is also an upper limit on the LSE cutoff at NNLO. There are some cutoffs, e.g., $\Lambda \approx 1250$ and $2300$ MeV, where the ${}^3$P$_2-{}^3$F$_2$ phase shift diverges.
At this order one subtraction is not enough to render the ${}^3$P$_2-{}^3$F$_2$ phase shifts stable with respect to cutoff.
We would have to add another, e.g. tensor, contact
term---a term that is nominally $O(P^4)$---in order to produce cutoff-independent results. This is analogous to the situation in the ${}^3$S$_1-{}^3$D$_1$ channel at NNLO~\cite{PVRA06A}, and can be dealt with in the subtractive approach provided an additional piece of experimental data is used as input~\cite{yangswavepaper}.

We then replaced the NNLO part of the DR TPE by SFR and again solved the LSE. Our results show
that in this case  the variation of $\alpha _{best}$ with $\Lambda$ is suppressed to less
than $5\%$. A comparison between the results obtained with the NLO TPE and this ``mixed" SFR TPE in
the $^{3}$P$_{2}-^{3}$F$_{2}$ channel implies that the mixed SFR NNLO TPE includes
more physics than the TPE at NLO.
Nevertheless, at larger cutoffs the repulsive NLO TPE
potential
in the $^1$P$_1$ and $^3$P$_0$ channels still leads to the same problem as observed in the DR NLO TPE, , and thus the allowed LSE cutoffs are still limited to $\Lambda=600$--$1000$ MeV.

Finally, we have the most success if we adopt the full SFR TPE at NNLO. In this case we obtain phase shifts which fit the Nijmegen ``data" reasonablely well and are (almost) independent of renormalization point for LSE cutoffs as high as $2500$ MeV.

To summarize our study, let us define necessary conditions for  
renormalization of potentials in the LS equation. The resulting phase shifts should:

\noindent (a) be (almost) independent of the cutoff $\Lambda$ in the LSE, \\
(b) fit the experimental analyses reasonably well, and be able to be improved order by order 
if the potential is obtained in a perturbative way, \\
(c) be (almost) independent of the point of renormalization.

Under these criteria,
our best-fits with dimensionally-regularized chiral potentials are limited to $\Lambda < 1$ GeV. Once cutoffs above this value are considered the Nijmegen phase shifts cannot be reproduced in certain channels at LO and NLO, while the instability of the ${}^3$P$_2-{}^3$F$_2$ phase shift violates condition (c) at NNLO. Therefore we conclude that chiral potentials should not be inserted into the LSE if cutoffs $\Lambda > \Lambda_{\chi {\rm SB}}$ are employed. If spectral-function regularization is employed the criteria (a) and (c) are satisfied for $\Lambda$ up to 2.5 GeV, but this is because another cutoff $\widetilde{\Lambda}$, which itself is below $\Lambda_{\chi {\rm SB}}$, has been used to ameliorate some of the features of the chiral potential that cause difficulties when it is inserted in the scattering equation.

{\bf Note added:}~After this paper was submitted for publication we became aware of Ref.~\cite{EG09}. In that paper an analytically solvable model for NN scattering in S waves is presented. An EFT which has the same long-distance physics as the model is then constructed. Epelbaum and Gegelia show that when the EFT integral equation is solved with cutoffs larger than the EFT breakdown scale it leads to erroneous conclusions about the dependence of phase shifts on the parameters that determine the long-distance physics. Although the model of Ref.~\cite{EG09} does not contain the singular potentials discussed in this work, the practical implications for chiral EFTs of NN scattering are in accord with the conclusions we have drawn here. 



\section*{Acknowledgments}
This work was performed in part under the
auspices of the U.~S.  Department of Energy, Office of Nuclear Physics,
under contract No. DE-FG02-93ER40756 with Ohio University. 
We thank the Ohio  Supercomputer Center (OSC) for the use of
their facilities under grant PHS206. D.R.P. acknowledges the hospitality of
the Centre for the Subatomic Structure of Matter at the University of Adelaide, where part of this work was done. 
All three authors are grateful for the hospitality of the Helmholtz Instit\"ut f\"ur Strahlen und Kernphysik, where this work
was completed. We thank Evgeny Epelbaum and Andreas Nogga for useful discussions. 


\clearpage


\begin{table}
\begin{tabular}{|l|llll|} \hline
    & $\alpha_{11}^{01}(^1P_1)$ & $\alpha_{11}^{10}(^3P_0)$ & $\alpha_{11}^{11}(^3P_1)$ &
$\alpha_{11}^{12}(^3P_2)$ \\
\hline\hline
NijmII     & 2.80 &  -2.47 & 1.53  & -0.28 \\
Reid93     & 2.74 &  -2.47 & 1.53  & -0.29 \\
Cd-Bonn    & 2.70 &  -2.45 & 1.51  & -0.29 \\
AV18       & 2.57 &  -2.41 & 1.44  & -0.30   \\
\hline\hline
\end{tabular}
\caption{The generalized P-wave scattering lengths $\alpha_{11}^{SJ}$
 as given in Ref.~\cite{vald} for the
NijmII and Reid93 potentials, together with ones extracted from the Cd-Bonn \cite{CDBONN}
and AV18 \cite{AV18} potentials. $\alpha_{11}^{SJ}$  is given in units fm$^{3}$.
}
\label{table-0}
\end{table}

\begin{table}
\begin{tabular}{|l|cccc|} \hline
    & $^1$P$_1$ & $^3$P$_0$ & $^3$P$_1$ & $^3$P$_2$ \\
\hline\hline
OPE    & U &  R & U  & R \\
NLO (DR)     & U &  U & R  & R \\
NNLO (DR)  & R &  R & R  & * \\
NLO (DR) + NNLO (SFR)       & U &  U & R  & R \\
NNLO (SFR)       & U &  U & R  & R \\
\hline\hline
\end{tabular}
\caption{Need for renormalization in different NN P-waves when the $\chi$ET potential is calculated to different orders. ``U" implies that the
potential does not need a subtraction to generate cutoff-independent phase shifts as $\Lambda \rightarrow \infty$ while ``R" means that a subtraction is required in that channel at that order. The * indicates that at NNLO the ${}^3$P$_2-{}^3$F$_2$ channel actually requires two subtraction to be made stable.}
\label{table-summary}
\end{table}

\clearpage


\noindent

\begin{figure}
\begin{center}
 \includegraphics[width=14cm]{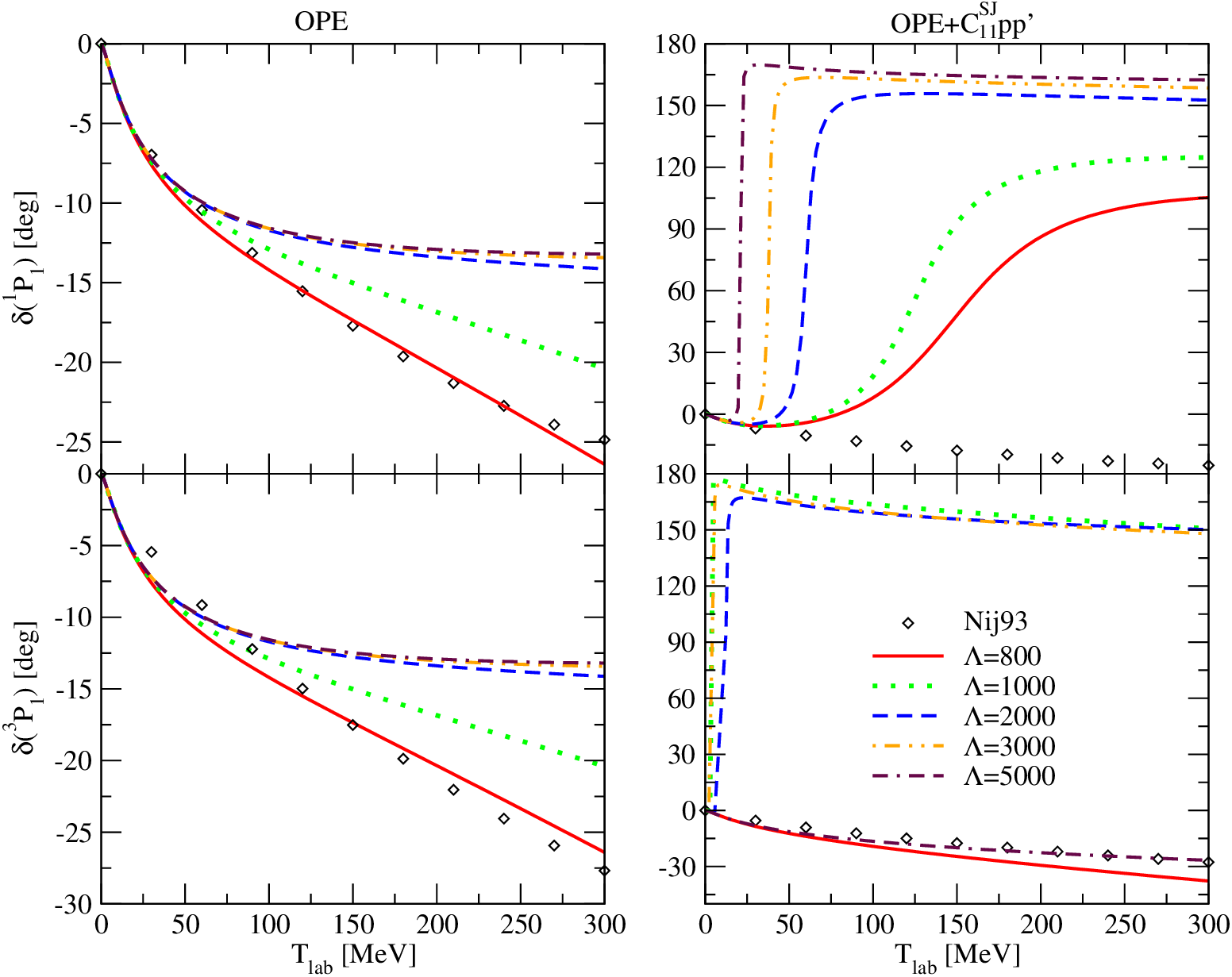}
\end{center}
\vspace{3mm}
\caption{(Color online) 
The NN phase shifts for the channels $^1$P$_1$ and $^3$P$_1$ as a function of
the laboratory kinetic energy for different cutoffs $\Lambda$ ranging from 0.8 to
5~GeV. In all cases the potential $v^{LR}=v_{1\pi}$ enters the LSE. The
panels on the left show the results without subtraction, whereas the panels on the
right show the results with one subtraction. The Nijmegen phase-shift
analysis~\protect\cite{nnonline} is indicated by the open diamonds.
\label{fig1}}
\end{figure}

\begin{figure}
\begin{center}
 \includegraphics[width=16cm]{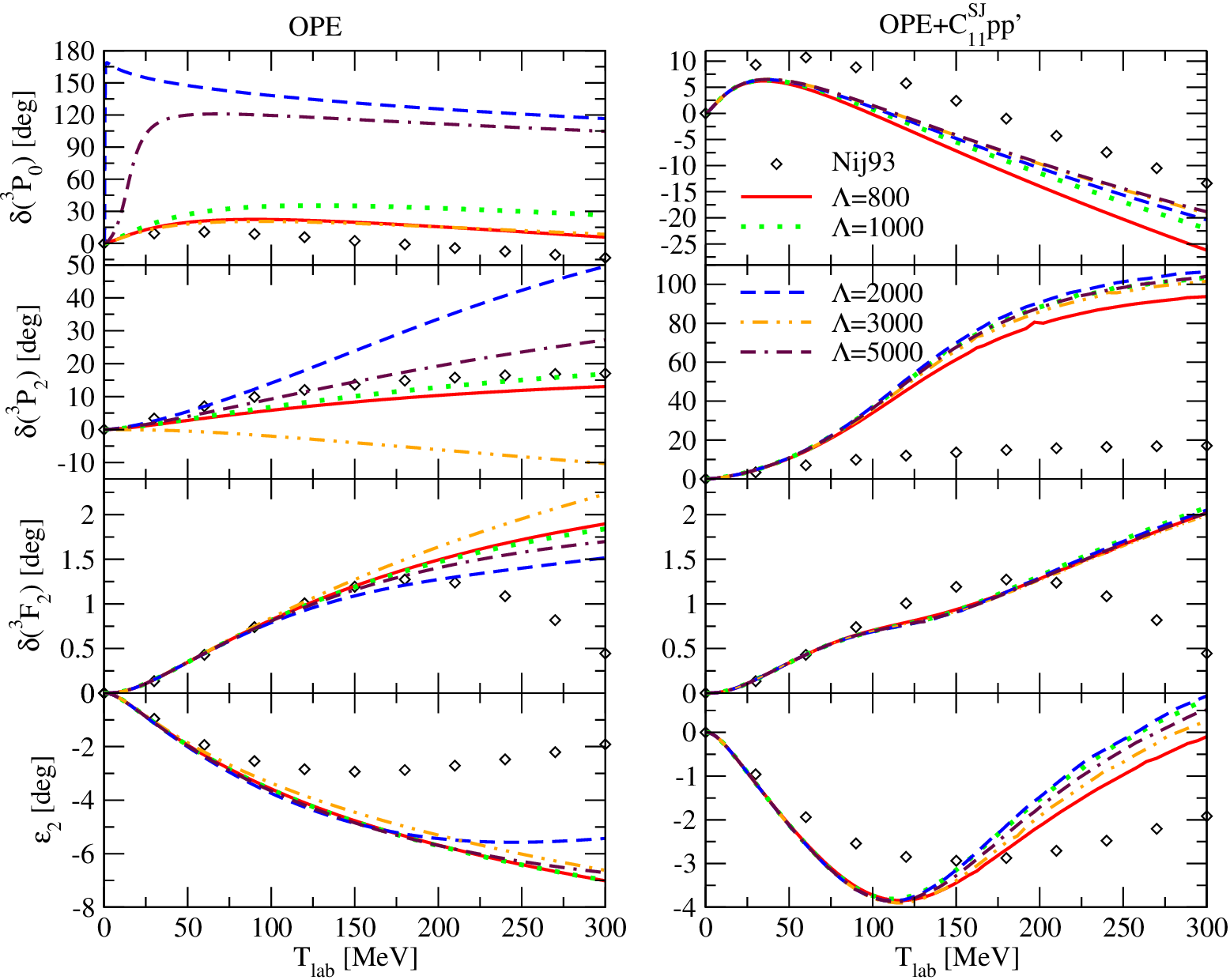}
\end{center}
\vspace{3mm}
\caption{(Color online)
The NN phase phase shifts for the channels $^3$P$_0$ and $^3$P$_2-^3$F$_2$ as
well as the mixing parameter $\varepsilon_2$  as a function of
the laboratory kinetic energy for different cutoffs $\Lambda$ ranging from 0.8 to
5~GeV. In all cases the potential $v^{LR}=v_{1\pi}$ enters the LSE. The
panels on the left show the results without subtraction, whereas the panels on the
right show the results with one subtraction. The Nijmegen phase-shift
analysis~\protect\cite{nnonline} is indicated by the open diamonds.
\label{fig2}}
\end{figure}

\begin{figure}
\begin{center}
 \includegraphics[width=16cm]{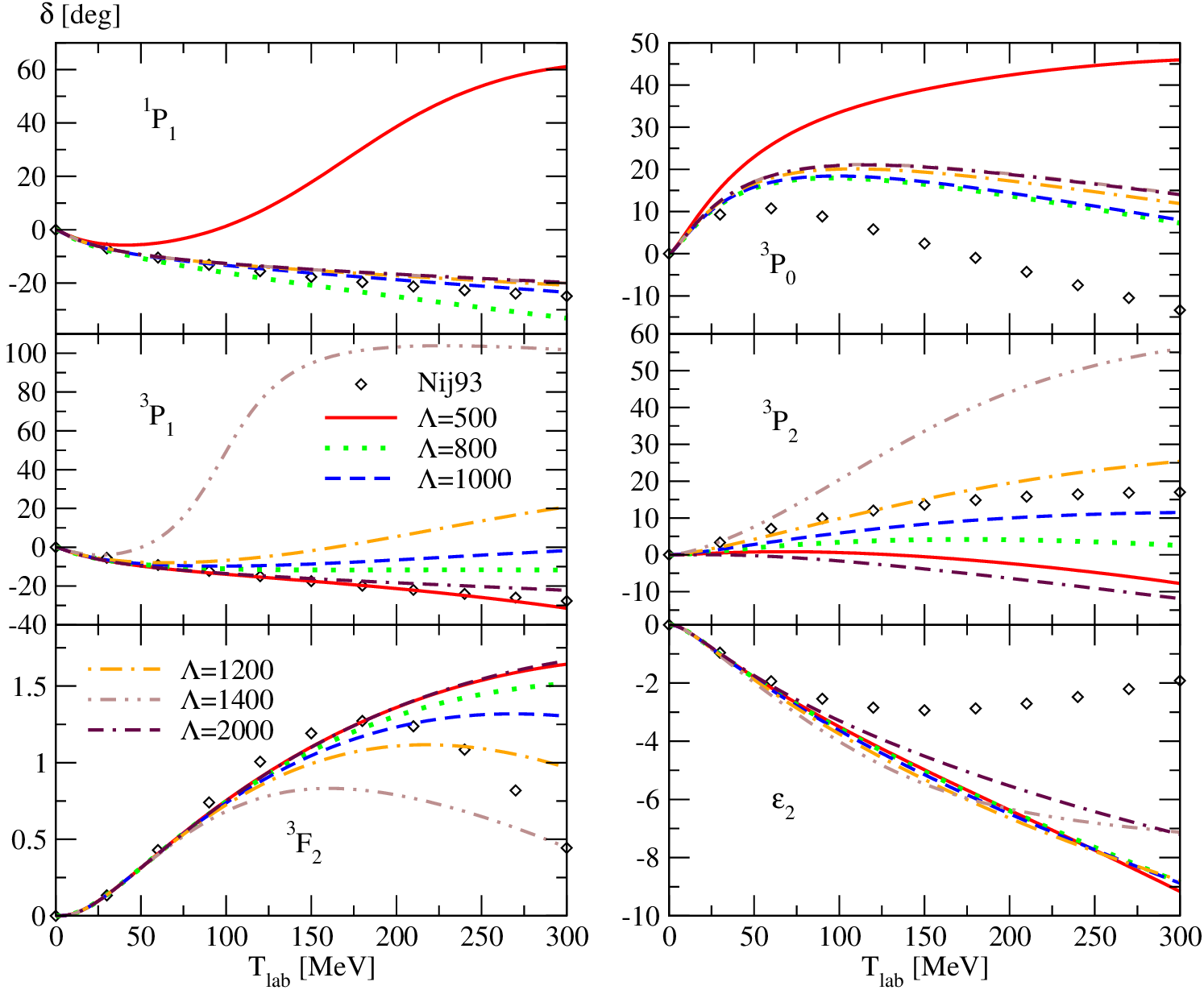}
\end{center}
\vspace{3mm}
\caption{(Color online)
The un-renormalized  NN P-wave phase shifts as a function of the laboratory kinetic
energy resulting from the dimensionally regularized TPE in NLO. The phase shifts are
shown for cutoffs $\Lambda$ ranging from 0.5 to 2~GeV.
The Nijmegen phase-shift
analysis~\protect\cite{nnonline} is indicated by the open diamonds.
\label{fig3}}
\end{figure}

\begin{figure}
\begin{center}
 \includegraphics[width=16cm]{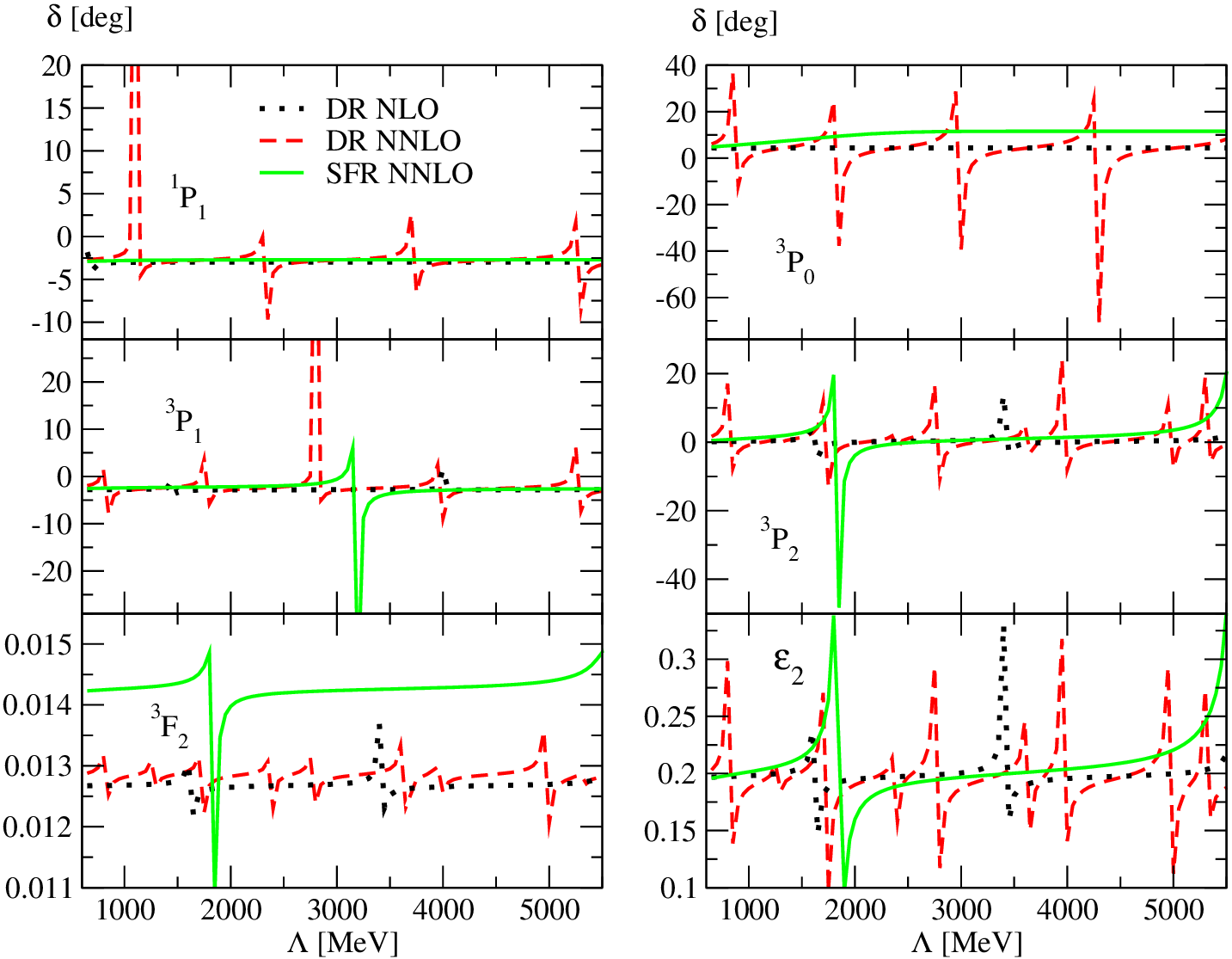}
\end{center}
\vspace{3mm}
\caption{(Color online)
The un-renormalized NN P-wave phase shifts at $T_{lab}=10$ MeV as a function of the cutoff for three different two-pion-exchange potentials: DR NLO, black dotted line; DR NNLO, red dashed line; SFR NNLO, solid green line. In each case $v^{LR}=v_{1 \pi} + v_{2 \pi}$. 
\label{figunrealla}}
\end{figure}

\begin{figure}
\begin{center}
 \includegraphics[width=16cm]{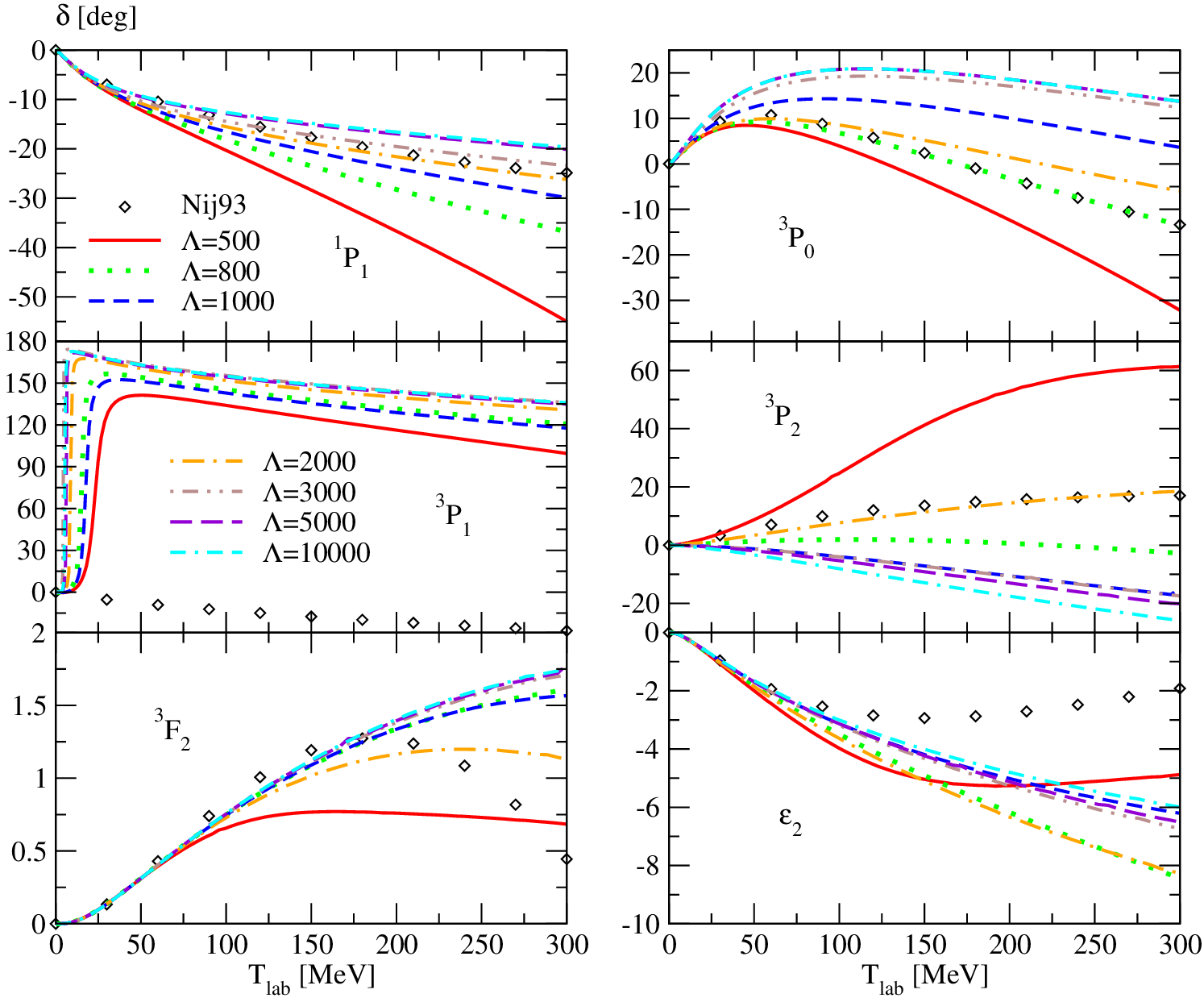}
\end{center}
\vspace{3mm}
\caption{(Color online)
The NN P-wave phase shifts  resulting from the use of OPE plus DR TPE at NLO with one
subtraction as a function of the laboratory kinetic energy. Here the cutoff range shown is 0.5--10~GeV. The input value of  $\alpha
_{11}^{SJ}$ is taken from Ref.~\protect\cite{vald}.
The Nijmegen phase-shift
analysis~\protect\cite{nnonline} is indicated by the open diamonds.
\label{fig4}}
\end{figure}

\begin{figure}
\begin{center}
 \includegraphics[width=16cm]{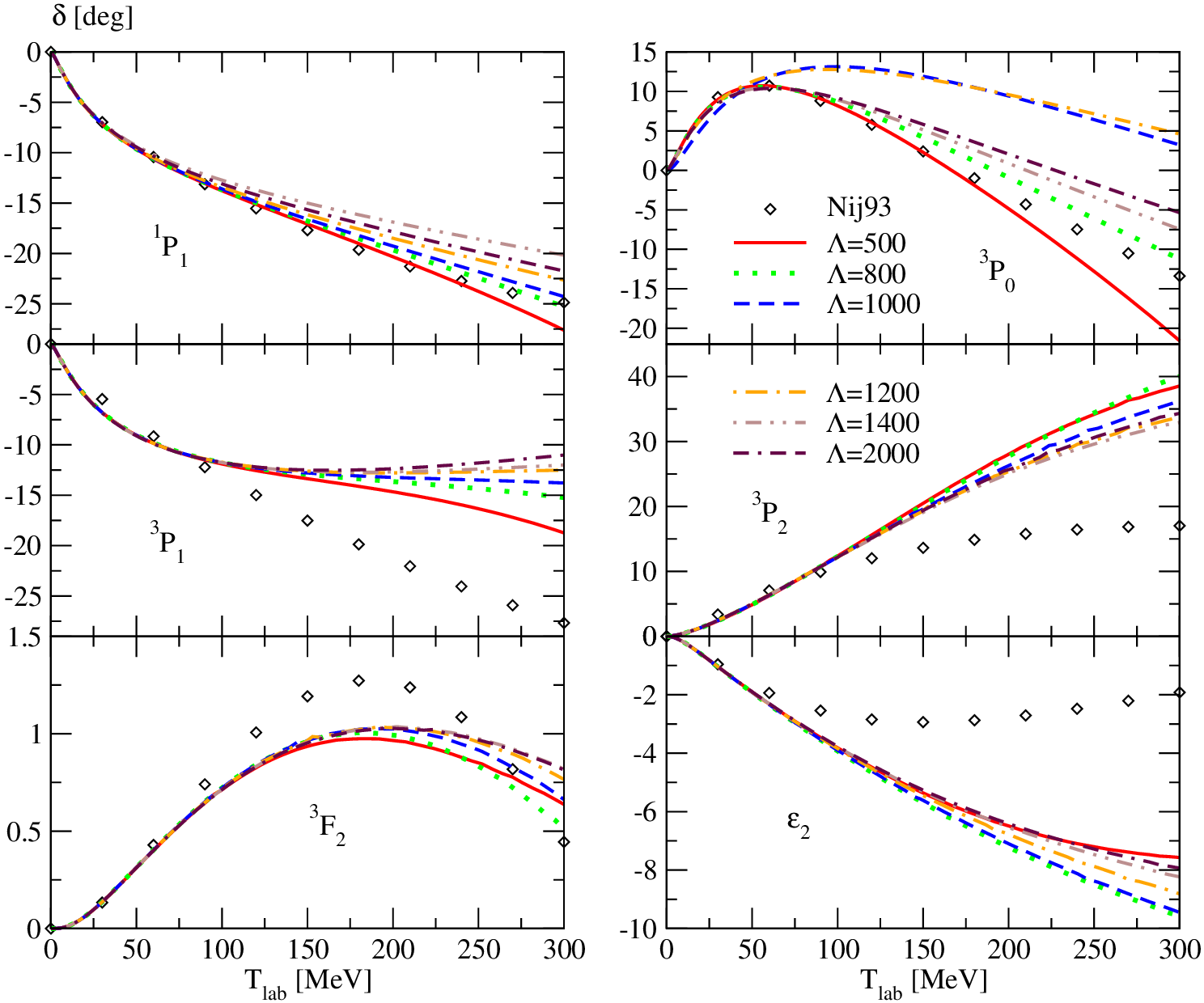}
\end{center}
\vspace{3mm}
\caption{(Color online)
The NN P-wave phase shifts as a function of the laboratory kinetic
energy that result from choosing $v^{LR}=v_{1 \pi} + v_{2 \pi}$, with the latter computed using DR TPE in NLO. Here 
the generalized scattering lengths $\alpha_{11}^{SJ}$ are adjusted  to give the best
fit in the region $T_{lab} < 100$~MeV. 
The Nijmegen phase-shift analysis~\protect\cite{nnonline} is indicated by the open diamonds.
\label{fig5}}
\end{figure}

\clearpage
\begin{figure}
\begin{center}
 \includegraphics[width=16cm]{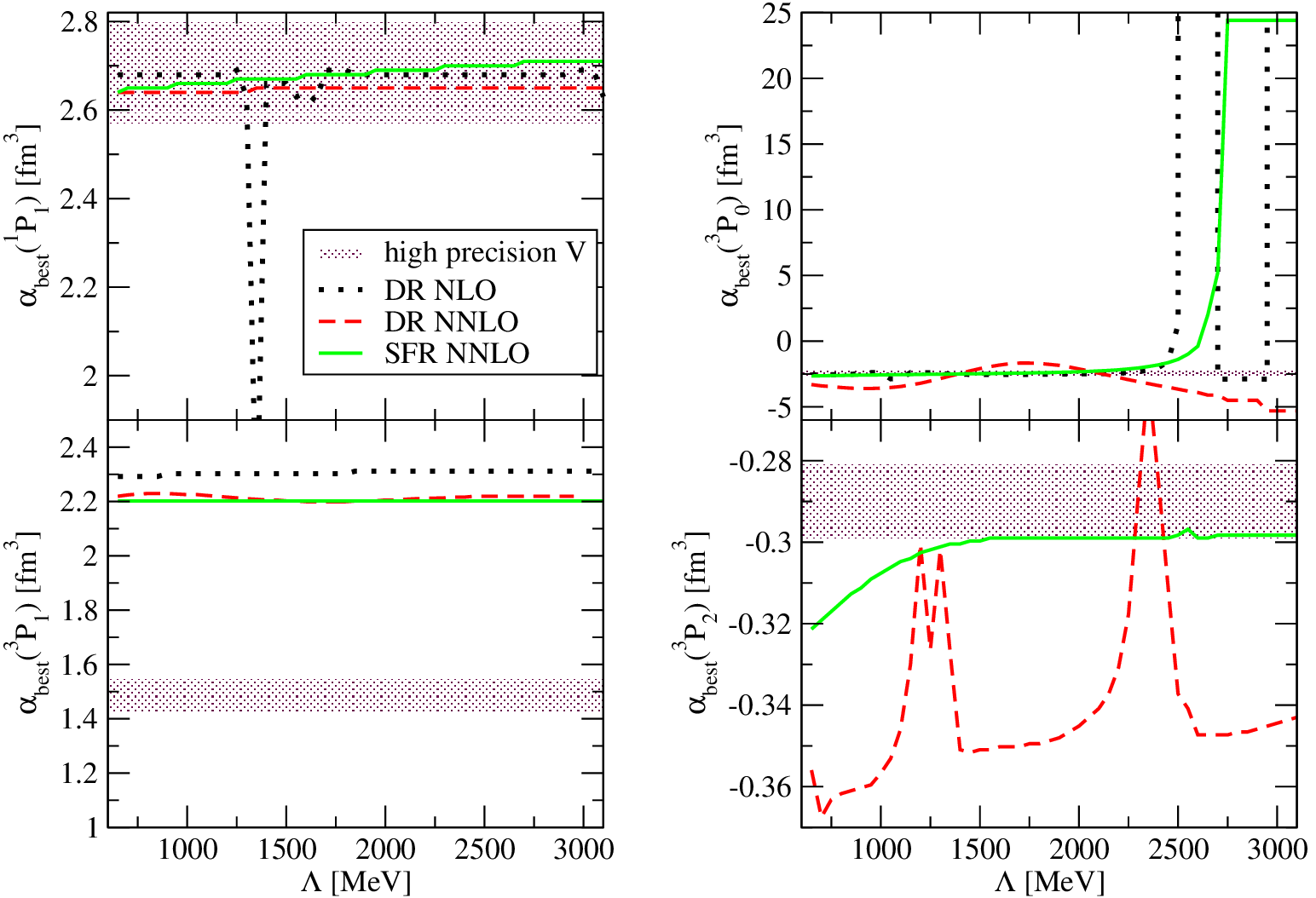}
\end{center}
\vspace{3mm}
\caption{(Color online)
$\alpha_{best}$ versus $\Lambda$ with the long-range potential chosen to be one-pion exchange plus DR NLO (black dotted line), DR NNLO (red dashed line) or SFR NNLO (green line). Here $\alpha_{best}$ has been adjusted at each cutoff to give the best
fit to the Nijmegen phase-shift analysis in the region $T_{lab} < 100$~MeV. The shaded region represents the range of the generalized scattering lengths extracted from the
`high-precision' potentials listed in Table I. 
\label{figalla}}
\end{figure}

\begin{figure}
\begin{center}
 \includegraphics[width=16cm]{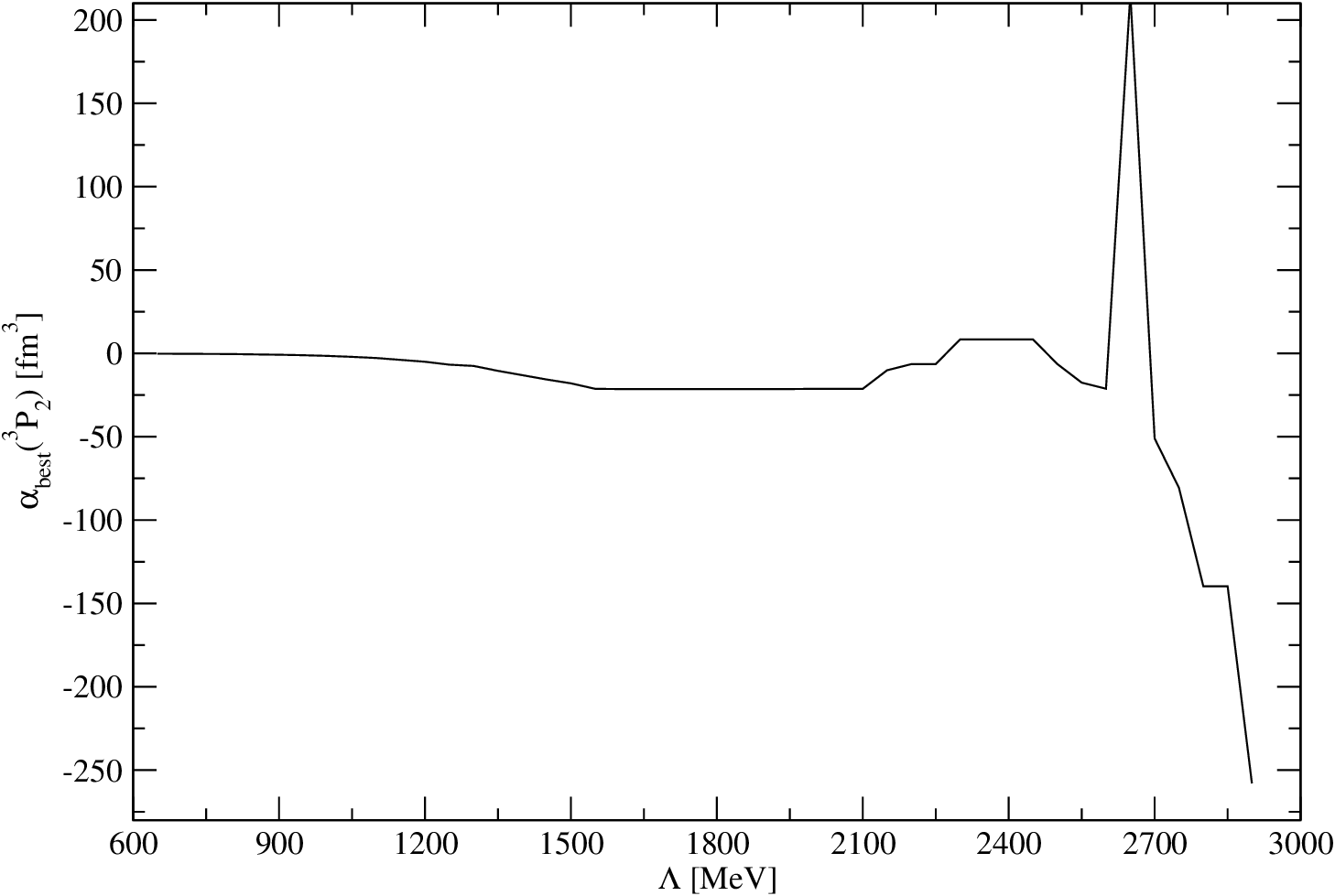}
\end{center}
\vspace{3mm}
\caption{
$\alpha_{best}$($^3P_2$) versus $\Lambda$ for $v^{LR}=v_{1 \pi} + v_{2 \pi}$, with $v_{2 \pi}$ computed at NLO using DR. Here $\alpha_{best}$ was adjusted at each cutoff to give the best fit to the Nijmegen ${}^3$P$_2$ phase shift
in the region $T_{lab} < 100$~MeV.
\label{figalpha3p2}}
\end{figure}

\begin{figure}
\begin{center}
 \includegraphics[width=16cm]{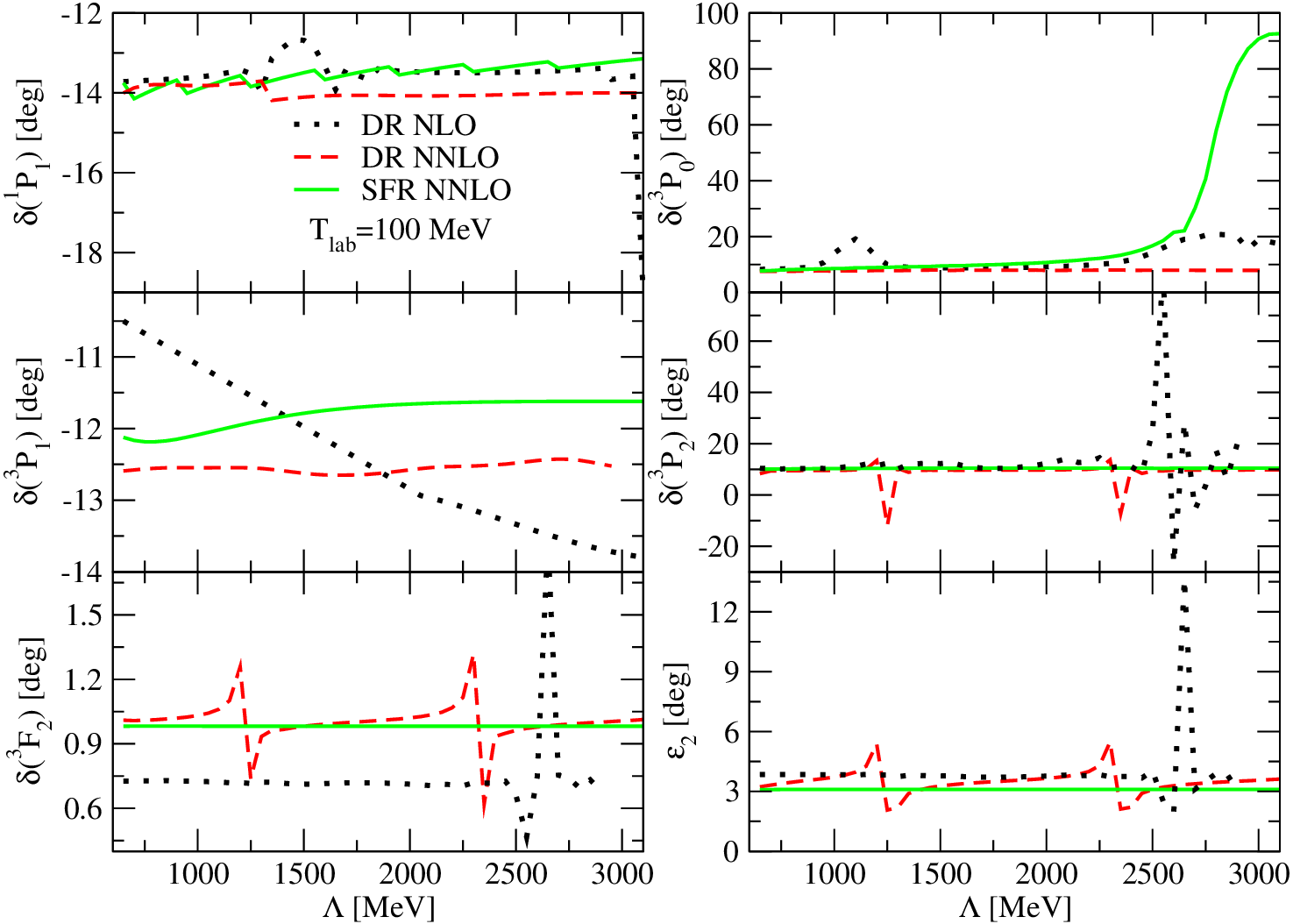}
\end{center}
\vspace{3mm}
\caption{(Color online)
The renormalized NN P-wave phase shifts at $T_{lab}=100$ MeV as a function of cutoff for three different $v_{2 \pi}$, with $v^{LR}=v_{1 \pi} + v_{2 \pi}$: DR NLO (black dotted line), DR NNLO (red dashed line), SFR NNLO (green solid line). 
The generalized scattering lengths $\alpha_{best}$ were adjusted at each cutoff to give the best
fit in the region $T_{lab} < 100$~MeV (see Figs.~\ref{figalla} and \ref{figalpha3p2}).
\label{figallbest100}}
\end{figure}

\begin{figure}
\begin{center}
 \includegraphics[width=16cm]{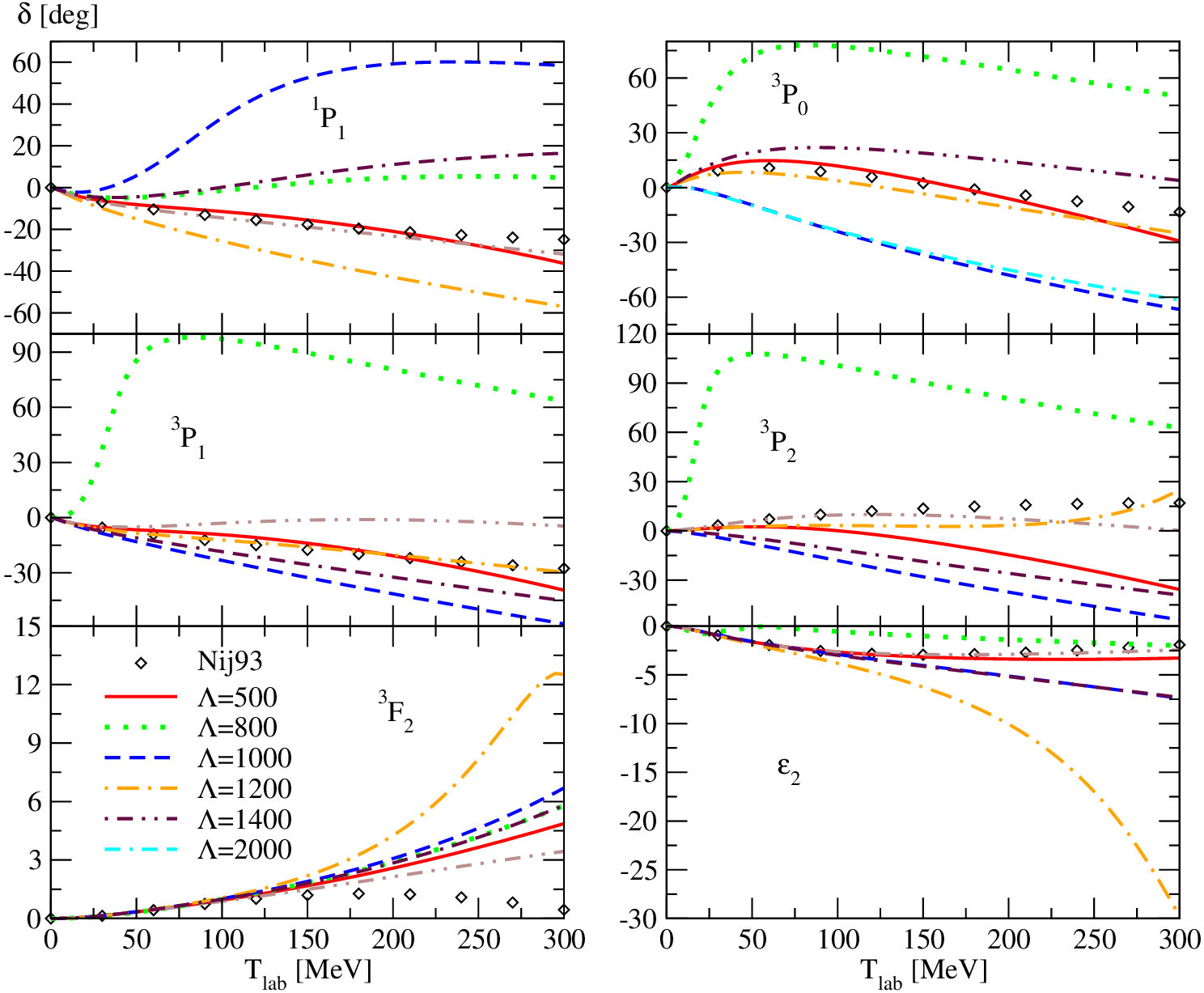}
\end{center}
\vspace{3mm}
\caption{(Color online)
The un-renormalized  NN P-wave phase shifts as a function of the laboratory kinetic
energy that result from a long-range potential of OPE plus dimensionally regularized TPE at NNLO. The phase shifts are
shown for cutoffs $\Lambda$ ranging from 0.5 to 2~GeV.
The Nijmegen phase-shift
analysis~\protect\cite{nnonline} is indicated by the open diamonds.
\label{fig6}}
\end{figure}

\begin{figure}
\begin{center}
 \includegraphics[width=16cm]{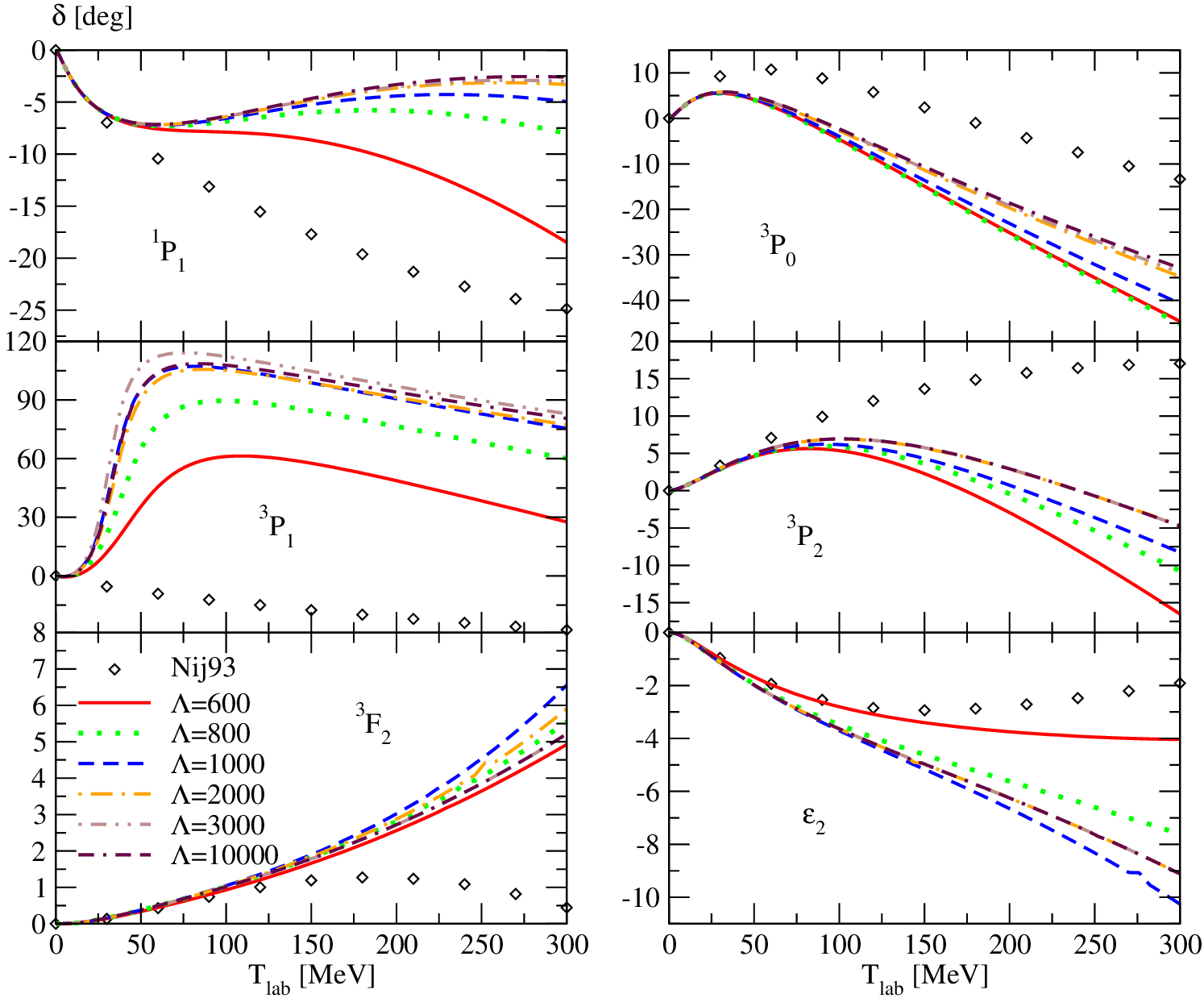}
\end{center}
\vspace{3mm}
\caption{(Color online)
The NN P-wave phase shifts as a function of the laboratory kinetic
energy that result from the use of NNLO DR TPE and one
subtraction. Here the cutoff range shown is 0.6--10~GeV. The input value of  $\alpha
_{11}^{SJ}$ is taken from Ref.~\protect\cite{vald}.
The Nijmegen phase-shift
analysis~\protect\cite{nnonline} is indicated by the open diamonds.
\label{fig7}}
\end{figure}

\begin{figure}
\begin{center}
 \includegraphics[width=16cm]{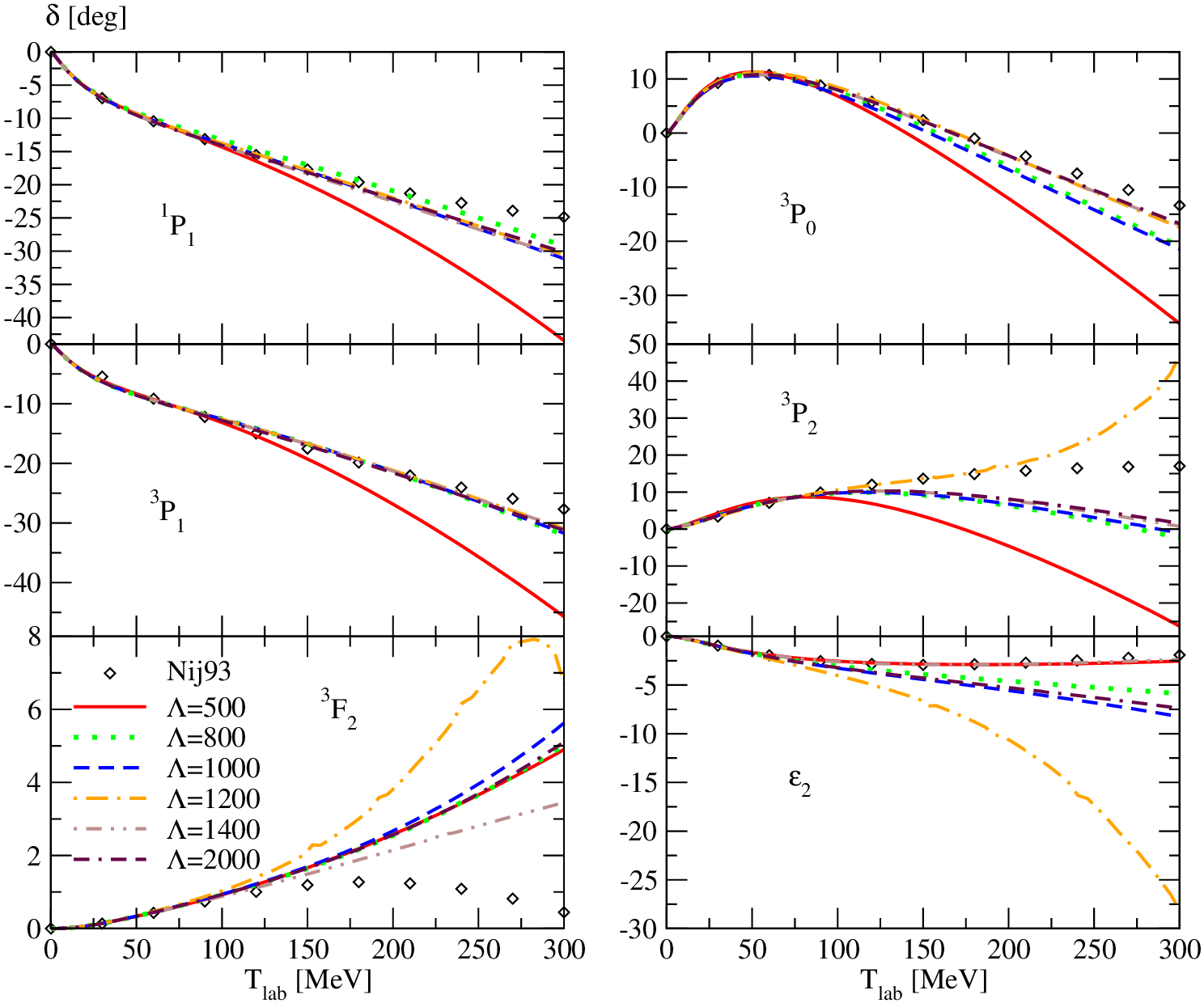}
\end{center}
\vspace{3mm}
\caption{(Color online)
The NN P-wave phase shifts as a function of the laboratory kinetic
energy  that result from choosing $v^{LR}=v_{1 \pi} + v_{2 \pi}$, with the latter chosen to be the DR TPE at NNLO, and implementing one subtraction. Here 
the generalized scattering lengths $\alpha_{11}^{SJ}$ are adjusted  to give the best
fit in the region $T_{lab} < 100$~MeV. 
The Nijmegen phase-shift
analysis~\protect\cite{nnonline} is indicated by the open diamonds.
\label{fig8}}
\end{figure}

\begin{figure}
\begin{center}
 \includegraphics[width=16cm]{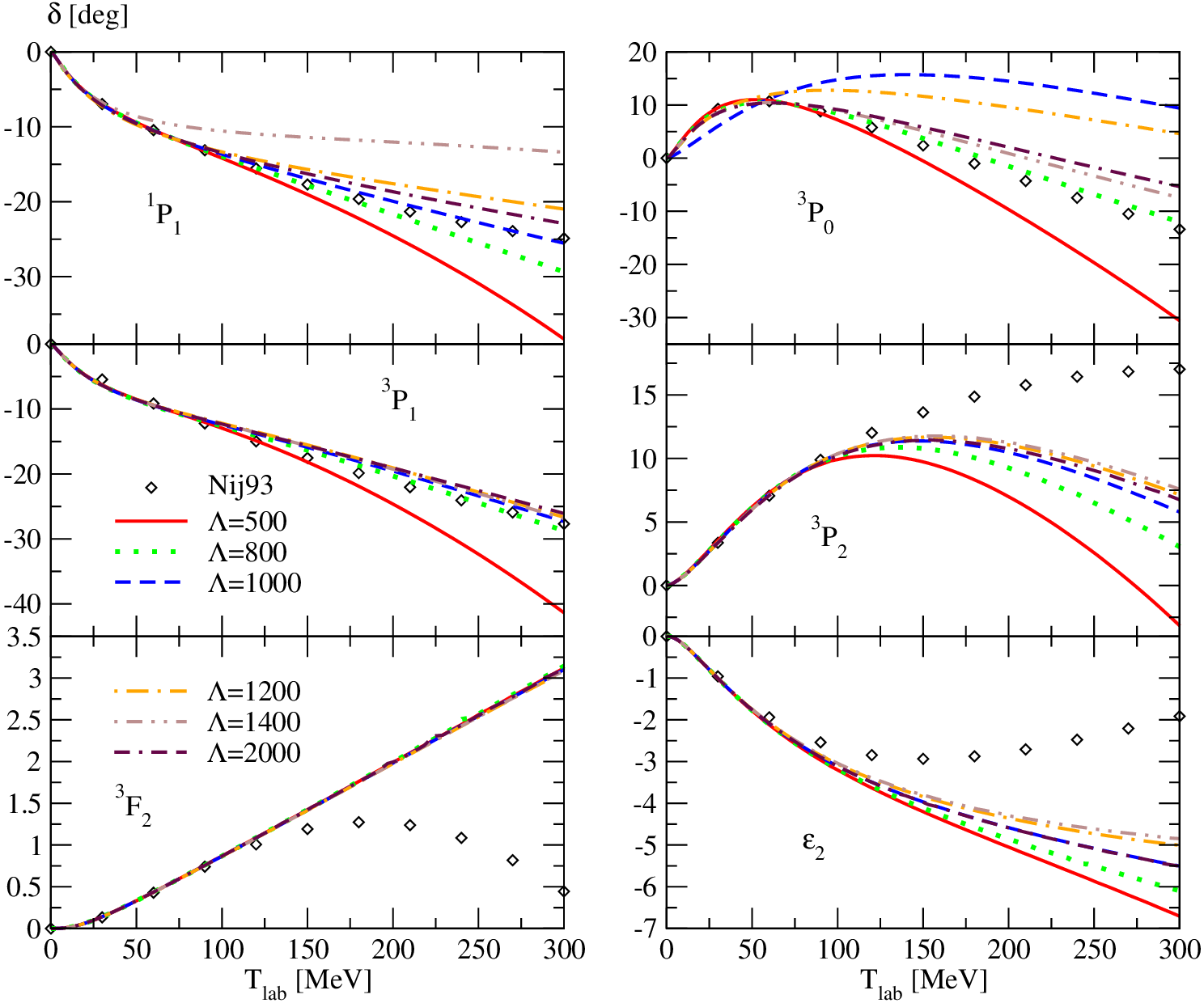}
\end{center}
\vspace{3mm}
\caption{(Color online)
The renormalized NN P-wave phase shifts as a function of the laboratory kinetic
energy  resulting when $v^{LR}$ is chosen to be the OPE plus TPE in NNLO, where
SFR is employed for the $O(Q^{3})$ part  of TPE. Here 
the generalized scattering lengths $\alpha_{11}^{SJ}$ are adjusted  to give the best
fit in the region $T_{lab} < 100$~MeV. 
The Nijmegen phase-shift
analysis~\protect\cite{nnonline} is indicated by the open diamonds.
\label{fig11}}
\end{figure}

\begin{figure}
\begin{center}
 \includegraphics[width=16cm]{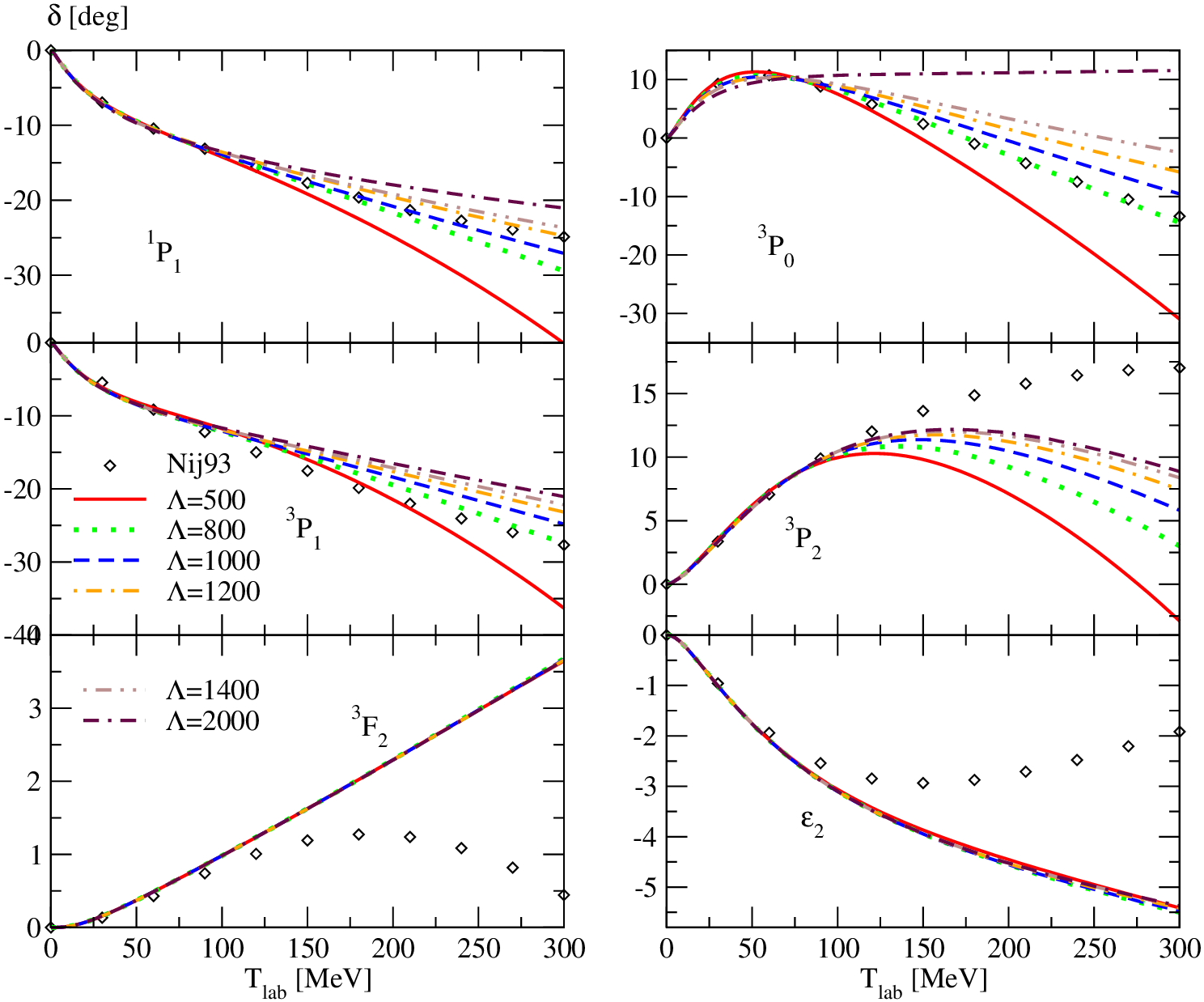}
\end{center}
\vspace{3mm}
\caption{(Color online)
The NN P-wave phase shifts as a function of the laboratory kinetic
energy  that result from choosing $v^{LR}=v_{1 \pi} + v_{2 \pi}$, with the latter chosen to be the SFR TPE up to NNLO, and implementing one subtraction. Here 
the generalized scattering lengths $\alpha_{11}^{SJ}$ are adjusted  to give the best
fit in the region $T_{lab} < 100$~MeV. 
The Nijmegen phase-shift
analysis~\protect\cite{nnonline} is indicated by the open diamonds.
\label{figfullbest}}
\end{figure}

\end{document}